# Exploring the relationship between journals indexed from a country and its research output: *An empirical investigation*

**Vivek Kumar Singh[a,1], Prashasti Singh[a], Ashraf Uddin[b], Parveen Arora[c], Sujit Bhattacharya[d]**

[a] Department of Computer Science, Banaras Hindu University, Varanasi, India.
[b] Department of Computer Science, AIUB, Dhaka, Bangladesh.
[c] Department of Science and Technology, Govt of India, New Delhi, India.
[d] CSIR- NIScPR, New Delhi, India.

**Abstract:** Scientific journals are currently the primary medium used by researchers to report their research findings. The transformation of print journals into e-journals during the last two decades has not only simplified the process of submissions to journals but has also increased their access across the world. It is well-known that there are significant differences in the total number of journals published from different countries. It is, however, not very concretely known whether the lack of appropriate number of publication venues in a country (including in one or more subject areas) may inhibit its publication propensity in one way or other. This article, therefore, attempts to explore the relationship between the number of journals published from a country and its research output.

Scopus database is used as reference database and the master journal list of Scopus is analysed to identify number of journals published from 50 selected countries, that have significant volume of research output. The publication data for the countries is obtained from Scopus. The following major relationships are analysed: (a) number of journals from a country and its research output, (b) growth rate of journals and research output for different countries, (c) global share of journals and research output for different countries, and (d) subject area-wise number of journals and research output in that subject area for different countries.

The analytical results show that for majority of the countries, the number of journals is positively correlated to their research output volume. A similar relationship is also observed in the subject area-wise analysis, confirming existence of the positive correlations between number of journals in a subject area and the research output in that subject area. However, several countries do not fully conform to the observed relationship, indicating that there are several other factors driving the research output of a country. The study, at the end, presents a discussion of the analytical outcomes and provides implications for policy perspectives for different countries.

---

[1] Corresponding author. Email: vivek@bhu.ac.in

**Introduction**

Scientific competency has become an important determinant for creating wealth and economic growth, as new technologies are increasingly science-based and draws from cross-disciplinary scientific fields. This has created competition among nations to develop a strong research 'niche' particularly in newly emerging areas. Developing a research ecosystem and infrastructure to support this ecosystem has thus become a priority in developed as well as emerging and developing economies. Journals are seen as a key scientific infrastructure of a country. Robert Merton (Merton, 1963) underscored the primary purpose of journals in addressing the contested claims for 'priorities' in research discoveries. He highlighted the key role of journals in settling disputes, and showed that the research publications in journals led to scientific disputes dropping to 72% in the 18th century, in the latter half of the 19th century it dropped to 59% and by 33% in the first half of the 20th century. The increasing acceptance of journals in the scientific community can be postulated happening due to various factors such as the peer review process becoming more institutionalized, frequency and fast dissemination of papers, journals availability in different disciplines/sub-disciplines and in newly emerging areas among others.

Journals have thus become the key communication channel of science and they have played this role right from the first journal published way back in 1665, named Philosophical Transaction of the Royal Society. Since then, thousands of journals have come into existence over a period of more than 250 years. The exponential growth of research papers in journals (Price, 1961) and an estimate by Jinha (2010) that 50 million research articles were in existence till 2009, are among the various indications that highlight that the scientific enterprise is intrinsically linked with journals. There are now different varieties of journals, some published and managed by professional societies, a large number by major commercial publishing houses, and some by higher education institutions. While majority of the journals are highly specialized- publishing mainly into a specific subject area, several others (such as Nature, Science) publish articles and scientific papers across a wide range of scientific fields. Journals are also often classified as national or international, depending on their overall character. However, with the emergence of new era of e-publishing, such differentiation in journals is becoming obsolete. The new e-publishing paradigm has also reduced different kinds of barriers related to article submission in journals and their dissemination to the readers.

The availability of large number of journals has perhaps motivated creation of multiple scholarly databases, which were initially used to measure and understand citation links. Over a period of last 50 years, several scholarly databases have come up. The Web of Science, Scopus, Google Scholar, Dimensions databases are some popular examples. There are also some subject specific databases such as Agricola, Mathsci, Pubmed, Inspec, dblp etc. These databases index journals published from different countries in different subjects and languages. Owing to being a repository of research metadata, these databases are now also being used for research evaluation exercises at different levels- individuals, institutions or countries. Scopus and Web of Science are the two most popularly used databases for such purposes. However, some studies (such as Singh et al., 2020; Mongeon & Paul-Hus, 2016; Singh et al., 2021a) have shown that the different databases have varied coverage of journals and, therefore, research evaluation exercises that use different databases often produce different outcomes. For example, Singh et al. (2020) have shown that India ranks at 11$^{th}$, 5$^{th}$ and 7$^{th}$ in global research output in Web of Science, Scopus and Dimensions, respectively, for the period 2010-19. The varied coverage of journals in databases is found to be the main reason for such variations. In this context, it is equally important to analyse and understand what impact the journal indexing by a database may have on the measured research output of a country, i.e., whether countries having higher number of journals indexed show higher research output.

Over the last few years, these databases have tried to expand their coverage by indexing a greater number of journals, more so from developing countries. This has indirectly impacted the numbers and suddenly the research output from some countries increased manifold. Some studies (such as Basu, 2010) have even shown for specific countries that the growth of national outputs was actually a matter of increased coverage of the concerned database. Many other studies (Leta, 2011; Collazo-Reyes, 2014; Erfanmanesh, Tahira & Abrizah, 2017) have pointed out towards this phenomenon in different settings, but often attributing the growth to several other factors operating together. It is in this context that this article attempts to systematically analyse the relationship between journals indexed from different countries and their research output. The Scopus Master Journal List[2] (updated till June 2020) is used for extracting the data for journals published from different countries, and the publication records for those countries are obtained from Scopus database[3]. The data is analysed to identify relationship between the number of journals indexed from a country and its research output, over a period of 15 years (2005-2019).

Examining this issue and framing the research problem, it was found that inspite of the high concentration by publishers dominantly coming from a few North economies, the affiliation of journals shows a wide dispersion. As visible in the two dominant databases, Web of Science and Scopus, one finds that the journals from many developing and emerging economies are now increasingly being indexed in these databases. Many studies in particular has pointed out the 'home bias' or 'home advantage' especially observed in citation influence (Tahamtan, Afshar & Ahamdzadeh, 2016; Gingras & Khelfaoui, 2018). Scholars have also examined this issue with journals published from a country as an indicator for accessing whether it influences research productivity (Meo et al., 2013; Mueller, 2016). The hypothesis that is built from this indicator is that the number of journals of a country may have an impact on its measured research output volume i.e., it is based on the 'home advantage' thesis. The influential studies in this direction are highlighted in the literature review. However, as we see later, the subject area-wise patterns in the relationship, the impact of advent of e-journals and the overall long-term trend has not been captured in earlier studies. Therefore, more empirical and analytical studies are required to understand the relationship between number of journals published from a country and its research output. The present paper addresses these issues and thus attempts to fill some important research gaps.

---

[2] https://www.elsevier.com/?a=91122
[3] www.scopus.com

Though, some previous studies (such as Basu, 2010; Leta, 2011; Najari & Yousefvand, 2013; Collazo-Reyes, 2014; Bhattacharya et al., 2014; Erfanmanesh, Tahira & Abrizah, 2017, Moed, Markusova & Akoev,) have addressed one or more aspect of the research question- whether indexing of higher number of journals from a country in a database lead to a higher research output for that country (in the database). However, these studies took data for a shorter period, the data was for early e-publishing era, and no subject area-wise analysis was done. Moreover, varying evidence is obtained in different studies, with Basu (2010), Collazo-Reyes (2014) and Erfanmanesh, Tahira & Abrizah (2017) showing evidence of a linear relationship, whereas, Leta (2011), Najari & Yousefvand (2013) and Moed, Markusova & Akoev (2018) attributing the observed patterns to various other factors. This study, therefore, not only attempts to bridge this research gap by carrying out an updated, systematic and comprehensive analysis, but also tries to settle the dispute regarding the relationship observed differently in different studies. The present study is unique in following respects:

- *First*, it uses data for a larger time-period of 15 years (2005-19) for the analysis, which is able to appropriately capture all kinds of patterns. Using this large time-period also allows to observe whether the rate of growth of number of journals of different countries correlates with their rate of growth of research output.

- *Secondly*, the study not only tries to understand the relationship between number of journals and research output for various countries, but also analyses whether the rate of growth of number of journals correlate with the rate of growth of research output. Similarly, the growth of global share of journals and growth of global share of research output for different countries is also analysed.

- *Thirdly*, the study provides an updated analysis on the topic, which is very useful and relevant given that the scientific publishing has now largely transformed into e-publishing.

- *Fourth*, it also performs a subject area-wise analysis of the relationship between number of journals in a subject area from a country and research output in that subject area for the country.

- *Finally*, the study identifies countries with exceptional performance of higher research output despite having lower number of journals and probable factors driving that are discussed, along with relevant policy suggestions for different countries.

**Related Work**

Several previous studies have dwelt into the question of database-induced variations in research output, rank and global share of different countries and have found that use of a different database for sourcing data may produce different outcomes. For example, Singh et al. (2020) have shown that the volume, rank and global share of ten highly productive countries vary across the Web of Science, Scopus and Dimensions databases, with the same countries ranked differently in different databases. These variations are essentially observed due to different number of journals indexed by different databases. The differential indexing of databases also implies that for a given country, different databases may be indexing different amount of its home or national journals, which in turn affect the research output volume and rank of the country in global perspective. Another set of studies (Mongeon & Paul-Hus, 2016; Singh et al., 2021a) have compared the journal coverage of Web of Science, Scopus and Dimensions databases and have shown that different databases have varied journal coverage which in turn results in different countries being ranked in different order on research output in different databases.

Several other studies (such as Basu, 2010; Leta, 2011; Michels & Schmoch, 2012; Najari & Yousefvand, 2013; Collazo-Reyes, 2014; Bhattacharya et al. (2014); Erfanmanesh, Tahira & Abrizah, 2017, Moed, Markusova & Akoev, 2018) have focused their attention on directly or indirectly understanding the relationship between indexing of journals from a country and its research output. Data for specific countries was analysed in different studies. We now look at some of these most relevant studies.

Basu (2010) considered declining productivity of India during 1990s and hypothesized that it was actually due to decline in number of Indian journals being indexed in the respective databases. The SCIMAGO data was analysed for 90 countries over the period 1996-2006 and a linear relationship between the number of journals indexed and the number of papers published was found, for a majority of countries. Some countries, like France, Japan and China, showed a pattern of higher number of their papers packed into their home journals, which led to introduction of Journal Packing Density (JPD). China's atypical rise in productivity in 2007 was largely attributed to high JPD.

Leta (2011) investigated the cause of rise of Brazil's productivity during late 1990s and early 2000, focusing on the question whether it was attributed to a true penetration of Brazilian science in the international arena or it was simply a result of an increase in number of Brazilian journals being indexed in academic databases. The study found that the rise in graduate courses in Brazil since 1990s enforced a strict criterion of productivity, that led to the adoption of international scientific models. This, in turn provided impetus to publish research internationally, which led to the enhancement in productivity. Several other initiatives such as Scientific Electronic Library On-line also made Brazilian science visible, which in turn impacted productivity. The study concluded that the home journals of Brazil in world's scholarly databases played a marginal and not a dominant role in the growth of Brazil's productivity, rather other factors were found contributing more to the cause.

Michels & Schmoch (2012) attempted to find the reason behind the surge of articles in the Web of Science (WoS) database during 2000-2008 period. The question explored was whether there had been an actual rise in scientific work around the world or the huge rise in scholarly publications was simply a result of addition of journals in the database. It was found that the traditional journals covered by WoS had swiftly decreased and newer ones were added. The study observed that out of the 34% rise in article growth, 17% was contributed by the inclusion of old journals that had been published for a long time but were not included by the database so far. The study, however, did not directly study the relationship between journals indexed from a country and its research output.

Najari & Yousefvand (2013) focussed on the scientific productivity of Iran in medical sciences and its contribution in this field to the Middle East and the world. Scopus Data for the time period 2000-2011 was analysed and different scientometric indicators (such as self-citations, % age of cited articles, international collaboration etc.) were computed. It was observed that in the year 2011, Iran accounted for 32.77% publications from the Middle East and 1.57% from the world, and ranked 17th and 23rd among the 226 countries in terms of number of articles and citations, respectively. It was concluded that the exponential growth of the country's research output was mainly due to the improvement of quantity and quality of indexed journals, which also indicated improvements in the research system of Iran.

Collazo-Reyes (2014) discussed about the unusual growth of Latin American and the Caribbean (LA-C) journals in Web of Science (WoS) in merely a four years span (2006-2009), It was observed that this was due to a change in the WoS editorial policy instead of the growth in the LA-C scientific community. As a result of this, the Portuguese language paved its way to become the second scientific language, after English, for LA-C journals in WoS. Among the LA-C countries represented in WoS, it was observed that Brazil comprised the highest share of scholarly data in WoS, which was an outcome of a larger share of its papers indexed in its home journals i.e., a high Journal Packing Density (an average of 100 articles per volume). Brazil's papers indexed in local journals comprised 26% of its entire WoS production, however, the citations received was just 7.5%. Further, 89% of Brazilian papers in its local journals were contributed by Brazilian authors. The Scopus database also comprised a significant share of LA-C (Latin American-Caribbean) journals but with a steady pace of growth in the considered time period (2006-2009). For the rest of the LA-C countries that were represented in WoS, low levels of productivity and impact was observed. The study, thus found a relationship between number of journals indexed and research productivity for some countries in LA-C.

Bhattacharya et al. (2014) examined the causality behind India's relative decline in the late 1980's and the publication growth from 1995 onwards based on both the databases Web of Science and Scopus. Drawing from global publication activity, the paper argued that India, to a large extent, epitomizes the scientific activity in emerging and developing economies. The study identified the following determinants of growth: expansion of journals in the global databases and significant increase in the Indian journals; expansion of institutes involved in publishing; increase in international collaboration; significant research activity in newly emerging areas. The inclusion of Indian journals progressively in the two databases (for example in 2005 there was 50 and 164 journals indexed in SCIE-E and Scopus, respectively which increased to 105 and 362 journals in 2012). It was an important factor as the average number of home papers in some journals were to an extent of 50% of the overall articles in the journal.

Erfanmanesh, Tahira & Abrizah (2017) analyzed the qualitative and quantitative role of country journals in the scientific performance of a country. They addressed three aspects: overall publication success of a country, correlations between the number of journals indexed and the number of papers published, and relationship between the number of papers published and the quality of country journals indexed in Scopus. The study analysed the data for 102 countries in the time period of 2005-2014 as obtained from SCImago Journal and country rank (http://www.scimagojr.com). It was found that for the majority of the 102 countries, the publication success largely depended on the number of country journals indexed in Scopus database, the number of papers published in country journals as well as the quality of the country journals indexed (as measured by indicators such as h-index, SJR, CiteScore etc.) The study found that Scopus comprised of maximum journals from Western Europe (48.9%) and North America (27.7%) with UK and US in a dominant position.

Moed, Markusova & Akoev (2018) performed a trend analysis of Russia's scientific productivity in the two popular academic databases- Web of Science and Scopus. Russia launched the Project 5-100 in 2013 with an aim to set the share of Russia's research output to 2.4% of the global Research Output and at-least 5 universities of Russia to feature among the world's top 100 universities ranked according to popular global ranking by the year 2015. The paper, thus, tried to identify factors contributing to the massive increase of Russia's publications in Scopus in 2000-2016. It was concluded that the publication counts and growth rate of publications from Russian institutions was very much impacted by the choice and coverage of database. Not only indexing of more Russian journals in databases contributed to the growth, but there was also an increase in Russian publications in other journals.

While most of these previous studies have tried to analyse the relationship between the number of country journals indexed in a database and the scientific productivity of the country, they did not agree on the type and magnitude of the relationship between the two. For example, Basu (2010), Collazo-Reyes (2014), & Erfanmanesh, Tahira & Abrizah (2017) support existence of a linear relationship between number of journals indexed from a country and its research output; but studies by Najari & Yousefvand (2013) and Moed, Markusova & Akoev (2018) have indicated that improvement in quality of research of the countries was also a major factor, with Leta (2011) suggesting that number of journals indexed have only a marginal role on the research output. Therefore, the relationship between number of journals indexed and research output of countries needs a fresh look to settle the disputed previous understanding. A fresh and updated analysis is also required due to the fact that journal publishing has now transformed to e-publishing mode. With the e-publishing mode becoming the main approach, journals are becoming more international in nature, and the barriers for submissions in the journals from across the world as well as for access to articles published in them have diminished. Therefore, one may expect that the relationship between number of journals indexed from a country and its overall scientific productivity, may have weakened over time. Accordingly, a fresh study with up-to-date data is required. Further, none of the previous studies analysed whether these relationships are also seen for different subject areas, in the sense that whether countries having higher number of journals in a subject area also have significantly higher research output in that subject area. Therefore, it would be interesting to analyse whether the research strengths of countries (measured as output in specific subject area) are related to number of journals in the subject area indexed in the database. The present study, therefore, attempts to fill this research gap and present an up-to-date analysis of the relationship between number of journals from a country and its published research output. The study not only tries to bridge the research gaps but also helps arriving at a more concrete understanding of the relationship between number of journals indexed and research output volume for a country, both in overall terms and also at the level of different subject areas.

**Data & Methodology**

The data for analysis was obtained from Scopus database. *First of all*, the Scopus Master Journal List (updated till June 2020) was downloaded from the Scopus website[4]. The Master Journal List contained 5 worksheets, namely Scopus Source Titles, Serial Conference Proceedings with Profile, All Conference Proceedings, More information on Medline, and ASJC Classification Codes. Out of these 5 worksheets, we used first worksheet which contained list of 40,385 source entries for a total of 114 countries. The list comprised of 3 source types- Journal, Book Series, and Trade Journal. There were 38,045 records for Journals, 1,527 records for Book Series and 812 records for Trade Journal. We have only considered the source type Journal which resulted in processing of 38,045 journal records. For each journal record in the list, there was a total of 55 fields. These fields included Source record ID, Source title, Coverage, Print-ISSN, E-ISSN, Active or Inactive, Source Type, Publisher's Country/Territory etc. We processed this list to identify journals published from different countries by using the field- Publisher's Country/Territory. Since, some journals were found to be inactive for certain years, therefore we did a year-wise sampling of active journals. This means that for a given year (say 2010), all journals which are active are included for analysis. When sampling for the next year (2011), if some of the journals 'active' in 2010 become 'inactive', they are excluded from the 2011 list. This was done due to the fact that Scopus will record publications for active journals only in a given year. In this way, the journals published from 114 different countries were identified, out of which we selected 50 countries, with highest research output in Scopus, for detailed analysis. In addition to overall counts of journals, the subject-wise counts of journals for the 50 countries were also obtained. The 27 major subject areas of Scopus were used for the purpose.

*Secondly*, for the 50 selected countries, the total number of publications were obtained from Scopus database for the period 2005-2019 by using queries of the form: (PUBYEAR > 2004 AND PUBYEAR<2020) AND AFFILCOUNTRY("X"), where X was substituted by the name of the country. A total of 50 such queries were executed and the publication records of document types 'article' and 'review' were downloaded. Only 'article' and 'review' document types were selected since they are the main publication items in journals. It may be noted that the data period selected for analysis is the one when majority of the traditional journals transformed to e-journals, with some even discontinuing the print publication.

*Thirdly*, the publication counts for the 50 countries in their home journals (referring to journals published from that country) was obtained from Scopus database for the period 2005-2019. This was done by using the query of the form: SRCID ($Y_1$ OR $Y_2$ OR $Y_3$…..OR $Y_n$) AND AFFILCOUNTRY("X") AND ((PUBYEAR> 2004 AND PUBYEAR<2020)), where $Y_1$, $Y_2$, …… $Y_n$ were the unique ids of the n journals published from a given country X. These publication records were also limited only to document types 'article' and 'review'.

---

[4] www.scopus.com

*Fourth*, the publication counts for the 50 countries in the 27 major subject areas was obtained from Scopus database for the period 2005-2019. This was done in two steps, first by formulating a query of the form PUBYEAR > 2004 AND PUBYEAR < 2020 AND AFFILCOUNTRY("X"). Then results retrieved from the above query were further processed by the Analyse Search Results tab provided by Scopus and under that, the results retrieved were listed according to 27 Subject areas. Thus, the publication counts for 50 countries in 27 major subject areas of Scopus for 2005-2019 were obtained. In this case too, publication records were limited only to document types 'article' and 'review'.

The data obtained as above was analysed to identify the relationship between number of journals indexed and the research output volume for the 50 countries. **Figure 1** presents a schematic diagram of the methodology used for the analysis. The number of journals, total publications, publications in home journals, and publications in different subject areas were identified and different relationships were analysed. The growth rate of number of journals and publications for the period 2005-2019 was computed by calculating Compounded Annual Growth Rate (CAGR) for each of the 50 countries. Two CAGR values, namely $CAGR_J$ and $CAGR_P$, referring to journals (J) and publications (P), respectively, were computed. The CAGR is defined by the following expression:

$$\text{CAGR} = \left( \left( \frac{Vfinal}{Vbegin} \right)^{\frac{1}{t}} - 1 \right) * 100$$

where, for $CAGR_P$, V*final* is the number of publication records in the year 2019, V*begin* is the number of publication records in the year 2005, and t is the time period in years. In a similar way, the $CAGR_J$ was also computed for all the countries.

The correlations between different values, for the given time period, are computed for all the countries, by computing Pearson correlation coefficient. For example, the Pearson correlation coefficient between NCJ and TP data for a country was computed using following expression:

$$\text{Correl (NCJ, TP)} = \frac{\sum(NCJ - \overline{NCJ})(TP - \overline{TP})}{\sqrt{\sum(NCJ - \overline{NCJ})^2 (TP - \overline{TP})^2}}$$

where, NCJ is the number of journals from a country, and TP is its research output. Correlation between other values was also computed in a similar manner.

The growth rate of number of journals and number of publications for the whole world was also computed. Two ratios, X1 and X2 were computed thereafter, for all the countries, where X1 refers to ratio of home journals of a country divided by total journals of the world, and X2 refers to ratio of total papers of a country divided by total papers of the world. Next, the subject-specific number of journals and publication counts for the 50 countries and correlations between them were analysed. Finally, the ratio of research output in home journals to the total research output was computed for all the countries. The different data were processed by writing programs in Python and various analytical results obtained are presented in tables and figures. The major variables computed and analysed are listed below:

| Variable | Description |
|---|---|
| NCJ | Number of Journals from a country |
| TP | Total Research Papers from a country |
| TPCJ | Total Research Papers of a country in Home Journals |
| NCJ-S | Number of Journals in a subject-area |
| TP-S | Total Research Papers in a subject-area |
| $CAGR_J$ | Compounded Annual Growth Rate of Journals |
| $CAGR_P$ | Compounded Annual Growth Rate of Papers |
| X1 | Ratio of home journals of a country divided by total journals of the world |
| X2 | Ratio of total papers of a country divided by total papers of the world |

**Results**

The analytical results first present some observations with respect to number of journals from different countries and the rate of their growth. Thereafter, the research output volume of different countries is analysed, followed by correlation between the two, to identify the relationship between them. The global share of journals from different countries and the global research output share of the countries are analysed next. A subject-specific analysis of the relationship between number of journals and research output of different countries is presented later. This is then followed by the results for the ratio of research publications in home journals to the total research publication, for different countries.

*Number of journals from different countries*

The number of journals published from the selected 50 countries during the period 2005-2019, along with the total number of journals published from the whole world during this period, are presented in **Table 1**. It can be observed that a good number of journals are published from US and selected European countries (UK, Netherlands and Germany, to name particularly). Since, we have the journal data for a period of 15 years, we have also computed the growth in number of journals for different countries by computing the Compounded Annual Growth Rate of number of journals (denoted as $CAGR_J$). We can see that the countries like South Korea (11.58%), Iran (15.31%), Malaysia (15.78%), Portugal (13.05%), Egypt (10.4%), Romania (13.4%), Thailand (13.99%), Indonesia (21.69%), Chile (11.93%), Colombia (16.19%) and Serbia (12.89%) have all recorded $CAGR_J$ value of greater than 10%, indicating good amount of growth in number of journals indexed in Scopus from these countries. Countries like US (1.38%), China (2.12%), UK (2.64%), Germany (2.53%), France (0.63%), Canada (0.77%), Netherlands (1.91%), Belgium (1.47%), Finland (1.45%) and Hongkong (1.42%) are the ones to have low growth recorded in the number of journals published. Japan (-0.04%) and Israel (-2.90%) are the two countries showing decline in number of journals indexed in Scopus during this period.

**Figure 2** shows the trend of growth/ decline of number of journals for the 50 selected countries. It is seen that South Korea, Brazil, Russia, Iran, Switzerland, Malaysia, Portugal, Romania, Indonesia, Colombia and Serbia show clear pattern of continuous growth in number of journals indexed in Scopus. Countries like Japan and France show initial growth but then a decline in number of journals indexed. Israel shows a declining pattern in number of journals indexed. For a recent relative picture of number of journals published from different countries, **Figure 3** shows the total number of journals published from the 50 selected countries as in the year 2019. It is seen that US, UK, Netherlands, Germany and Switzerland have the highest number of journals published. In fact, US, UK, Netherlands and Germany, these four countries, taken together, alone account for about 63% of the total journals published in the world. Other countries have much lower number of journals published in comparison to these countries, though several countries are growing rapidly in terms of number of journals published and indexed in Scopus. The relative positions of different countries on the number of journals published is thus clearly understood from this figure.

*Research output from different countries*

The total number of publications for the 50 selected countries during 2005-2019 period are obtained from Scopus database. As indicated earlier, these counts are only for document types 'article' and 'review'. **Table 2** shows the publication counts along with the Compounded Annual Growth Rate of publications (denoted by $CAGR_P$). It is seen that US, China, UK and Germany are the leading countries in terms of research papers published in this period. In terms of rate of growth, it can be observed that countries like India (10.13%), Iran (16.27%), Malaysia (17.75%), Egypt (11.94%), Saudi Arabia (18.01%), Pakistan (15.87%), Indonesia (23.45%), Colombia (15.65%) and Serbia (11.11%) are the ones having high growth rate of number of publications. Countries like US (2.66%), UK (3.76%), Germany (3.12%), Japan (0.93%), France (3.05%), Canada (4.18%), Taiwan (3.43%), Israel (3.15%), Greece (3.48%) and Hungary (3.64%) have relatively low growth in the number of publications. However, many of these countries have a high volume of research output, and hence despite the growth rate being low or moderate, the absolute amount of research output is significant. For example, China has a $CAGR_P$ value of 9.65%, which is good growth rate, particularly given that its publication volume is already significantly large. Interestingly, as we will see in the next section, most of the countries having high $CAGR_J$ also have high $CAGR_P$, indicating that the rate of growth of journals from a country is related to rate of growth of publications of that country.

*Relationship between number of journals and research output*

In order to understand the nature of relationship between number of country journals indexed and its research output, first of all, correlation was computed between the two for all the 50 countries. For example, for the country US, Pearson correlation is computed between its number of journals during 2005 to 2019 and its research output during the period. This is then repeated with data for all the countries. **Table 3** presents the Pearson correlation coefficients for the 50 countries. It may be observed that the Pearson correlation coefficient for 32 out of 50 countries is higher than 0.9. Thus, except for some countries like Israel and Japan, majority of other countries have high Pearson Correlation Coefficient value, indicating a moderate to strong positive correlation between number of journals and research output of the countries. The average of all the correlation values for all 50 countries is 0.84. Most of the correlations (except for Japan and Israel) were found significant at 0.01 level (2-tailed). The high values of correlation coefficients can thus be taken as an indicator of positive relationship between number of journals and research output of a country.

To get further insight in the relationship between two variables, we tried to observe the relationship between $CAGR_J$ and $CAGR_P$ values too, for all the countries, i.e., whether the countries with high $CAGR_J$ have high $CAGR_P$ and those having low $CAGR_J$ have low $CAGR_P$. We observed many examples of such kind, such as Iran, Malaysia, Egypt, Indonesia, Colombia and Serbia, all of which had both $CAGR_J$ and $CAGR_P$ above 10%. Similarly, countries like US, UK, Germany, France, Canada, Japan, Israel, Finland are the prominent examples of countries having low values of both $CAGR_J$ and $CAGR_P$. Motivated by these observations, we plotted a scatter plot of $CAGR_J$ and $CAGR_P$ values of the 50 selected countries, as shown in **Figure 4**. Countries like Indonesia, Malaysia, Iran and Colombia etc. are the ones having both $CAGR_J$ and $CAGR_P$ in the high value quadrant. Similarly, a large number of countries are in the quadrant of low $CAGR_J$ and $CAGR_P$ values. Some exceptions are China, India, Saudi Arabia, and Pakistan, all of which have relatively moderate $CAGR_J$ value but higher $CAGR_P$ value. We tried a linear regression fit for the observed patterns and found that $R^2$ value is 0.693. A value of 0.693 for $R^2$ is usually considered a moderate fit. Thus, we can conclude that there is a moderate linear relationship between $CAGR_J$ and $CAGR_P$, with about 69% points being a good fit to the straight line.

In order to have the relationship between number of journals from a country and its number of publications analysed in yet another way, two ratios - X1 and X2 - are also computed. The value X1 represents the ratio of number of journals from a country to the total number of journals in the world, as indexed in Scopus for all the years under consideration. The value X2 represents the ratio of number of publications from a country to the total number of publications from the world, as indexed in Scopus for all the years under consideration. **Figure 5** plots the values of X1 and X2 for all the 50 selected countries for the period 2005-2019. The X1 and X2 patterns drawn in the figure show that for majority of the countries X1 and X2 have similar pattern, indicating that as the global share of its number of journals increase, the global share of its publications has also increased. Some notable exceptions are, however, there in form of countries like China, India, Malaysia, South Africa, Saudi Arabia, Iran, and Pakistan, all of which have higher growth in global research output share as compared to global journal share. Similarly, Switzerland and Romania are prominent examples of decrease in global research output share despite an increase in the global journal share.

*Subject area-wise analysis of number of journals and research output*

The relationship between number of journals and research output of the countries was also analysed for different subject areas. The objective was to see if countries having higher number of journals in a subject area get higher research output in that subject area. In other words, whether there exists a linear relationship between number of journals of a country in a given subject area and its research output in that area. For this purpose, the 27 major subject categories of Scopus were chosen as subject areas. For each subject area, the number of journals were identified for each country through a year-wise sampling, as used earlier (referred to as NCJ-S). Similarly, the research output in each subject area for each country was also obtained (referred to as TP-S). Thereafter for each country the correlation was obtained between year-wise values of NCJ-S and TP-S. **Figure 6** shows a matrix of Pearson correlation coefficient values between NCJ-S and TP-S for the 50 countries in the 27 subject areas. The 'green' color denotes values greater than '0' and 'red' color denotes values less than '0'. Intensity of the color denotes the strength of correlation. Some values are not available (marked by 'white' color). It is observed that the majority of the values are positive, indicating positive correlation between NCJ-S and TP-S. In fact, out of 1,350 correlation coefficients, 277 values are NA, 955 are greater than '0', and only 118 are negative. Among values greater than '0', 703 values are in the range 0.7 to 1.0, 131 values are in the range 0.5 to 0.7, and 121 values are between 0 to 0.5. Thus, a large majority of the correlation values between NCJ-S and TP-S show positive relationship between number of journals in a subject area from a country and its output in that subject area. In other words, those countries that have higher number of journals published in a subject area, also get higher number of publications in that subject area. It would be interesting to mention here certain examples. We can see that US and UK have good number of journals in Agricultural & Biological Sciences (AGRI) and Arts & Humanities (ARTS) and also good research output volume in these subject areas. Australia and Brazil have lesser number of journals in Chemical Engineering (CENG) and also low research output volume in this subject area. There are, however, also some exceptions, indicated by a negative correlation.

*Ratio of publications in home journals to total publications*

We analysed what proportion of research publications from a country appear in its home journals. This is done to see if countries with high growth of research output are actually getting higher proportion of its research papers appearing in home journals. The ratio of TPCJ and TP are computed for all the 50 countries over the period of 15 years. **Table 4** presents the ratio of publications in home journals (TPCJ) to the total publications (TP) for different countries. In general, for most of the countries the proportion of research output appearing in home journals have declined, except for some countries like Switzerland and Malaysia. For countries like US, UK, Germany, which has large number of home journals, the decline in publications in home journals is not that significant. Some other major countries that show a decline in proportion of papers in home journals are France, Poland, South Africa, Pakistan etc. Most of other countries either continued with almost similar proportion of papers in home journals or show a very slight decline over time. This is an indirect indication that their growth in research output involves factors beyond publications in nationally oriented journals. Few countries show an increase in the proportion of papers in home journals, such as Switzerland and Malaysia, both of which saw an increase in number of journals indexed. One interesting pattern worth observing is that of China. It is seen that China, which published 63% of its papers in home journals during 2005, shows a constant decline of proportion values, becoming just 17% in the year 2019. This is a clear indication that China's total publication growth is significantly higher as compared to its journal growth. In fact, China, has expanded its publication base significantly during this period. These results provide a general observation that majority of the countries have expanded their research publishing beyond home journals, since the proportion of papers published in home journals has decreased from majority of the countries.

**Discussion**

The analytical results about number of journals and research publications show interesting patterns. First, we discuss some observed trends about number of journals from different countries indexed in the Scopus database. It is observed that while US, UK, Germany and Netherlands continue to have the highest number of journals indexed in Scopus, several other countries have also witnessed a rapid increase in number of journals indexed. This growth in number of journals indexed is happening more in southeast Asian countries (like Indonesia, Thailand, South Korea), Western Asian countries (like Iran) and non-English speaking European countries (like Spain and Portugal). Thus, Scopus appears to be indexing more and more journals from a wider list of countries. For many countries, the number of journals indexed has multiplied manifolds during the 2005-19 period. Examples include South Korea (5 times), Iran (8 times), Malaysia (9 times), Brazil (4 times), Portugal (6 times), Romania (6 times), Indonesia (19 times), Colombia (9 times), Thailand (7 times) and Serbia (6 times). The above statistics provide a broad overview of the global journal landscape based on country affiliation. Causality behind the trend may be influenced by a large number of latent factors. The presence of big publishing house, reputed universities and research institutes and professional/academic societies, liberal endowment funds can be plausible factors for the skewness observed in journal affiliations country wise. The demand of inclusivity and addressing global audience could be the motivating factor behind journals from emerging and developing country journals getting included in the database. Addressing the bias of journals only from English speaking country may also be another reason behind the shift that is observed in journal inclusion. Therefore, a more in-depth study is required to arrive at a more informed understanding of the distribution of journal indexing from different countries.

We now discuss the observed relationship between number of journals indexed from a country and its research output. It is observed that the number of journals (NCJ) and research output (TP) from a large majority of countries, are found to be correlated well. Out of 50 countries, 32 have Pearson correlation coefficient greater than 0.9, indicating a strong positive correlation. These patterns thus indicate that countries having higher number of journals indexed do also have higher research output. A similar kind of positive association was also observed in the previous studies by Basu (2010), Collazo-Reyes (2014) and Erfanmanesh, Tahira & Abrizah (2017), all of which found a linear relationship between number of journals from a country and its research output. However, at the same time, we observe that some countries do not show a strong positive correlation, or at least the rate of growth of journals ($CAGR_J$) is not commensurate with the rate of growth of research output ($CAGR_P$). For example, Israel and Japan have the lowest correlation values. Similarly, as in 2019, China, which is $2^{nd}$ in world in terms of research output has a smaller number of journals indexed ($7^{th}$ rank according to number of journals indexed). Therefore, China's growth in research output cannot be attributed much to growth of its number of journals. Another example is India, which is ranked 6[th] in global research output, but has only limited number of journals indexed as in the year 2019 (11[th] rank according to number of journals indexed). Thus, in case of India too, other factors have also played a role in growth of research output. Netherlands, which has the 3[rd] highest number of journals indexed, ranks at 15[th] place in global research output. These examples thus support findings of Leta (2011), Najari & Yousefvand (2013) and Moed, Markusova & Akoev (2018), all of which indicate that growth of research output of a country has to be attributed to both quantity (growth in journal indexing) and quality (increase in internalization and quality of science) factors.

The analytical results also show that for a majority of countries, the rate of growth of research output ($CAGR_P$) correlates with the rate of growth of journals ($CAGR_J$). The results show a close to linear relationship between $CAGR_J$ and $CAGR_P$ (Figure 4), thus supporting findings about linear relationship by Basu (2010), Collazo-Reyes (2014) and Erfanmanesh, Tahira & Abrizah (2017) However, in this case too, there are several exceptions such as Saudi Arabia, Pakistan, China, India and South Africa, all of which have higher rate of growth of research output as compared to rate of growth of journals. Similarly, countries like South Korea, Thailand, Romania and Portugal register a high growth rate in number of journals but their research output growth is not in the same order. For these countries, the research output volume can be attributed to other factors (such as quality), as suggested by Leta (2011), Najari & Yousefvand (2013) and Moed, Markusova & Akoev (2018). The plots for X1 and X2 (Figure 5) also show interesting patterns of relationship. For majority of the countries, X1 and X2 follow similar trend, i.e., as the countries' global share of journals has increased, its global share of publications has also increased. However, again exceptions are observed in form of countries like China, India, Saudi Arabia, South Africa and Indonesia, all of which is growing higher in terms of global share of research output as compared to growth of global share of number of journals. These countries have improved their research output volume at a rate which is higher than rate of growth of journals. This indicates that they have also witnessed an overall improvement of standards of research publishing so as to get the publications accepted in the well-known international journals. The e-publishing era may also be acting as a facilitator for wider acceptability and dissemination of research conducted in these countries.

The subject area-wise analysis of the relationship between number of journals and research output of countries, also shows a kind of positive correlation for majority of the countries. It is observed that for majority of the countries the correlation coefficient computed is positive and strong (Figure 6). It can be seen that, out of 1,350 correlation coefficients, 955 are greater than '0', with 703 values in the range 0.7 to 1.0 and 131 values greater than 0.5. Thus, a large majority of the correlation values between NCJ-S and TP-S show positive relationship between number of journals in a subject area from a country and its output in that subject area. In other words, those countries that have higher number of journals published in a subject area, also get higher number of publications in that subject area. For example, US and UK have good number of journals in Agricultural & Biological Science (AGRI) and Arts & Humanities (ARTS) and also good research output volume in these subject areas. Australia and Brazil have lesser number of journals in Chemical Engineering (CENG) and also low research output volume in this subject area. However, in case of subject area specific patterns too, exceptions are observed, as seen in the 'red' colored values in the Figure 6.

In order to ascertain this attribution of increase in research output to increase in number of journals, one would need to analyse the pathways of the relationship between the two. Is the relationship happening due to a common underlying factor that is affecting both the variables positively is an important question that needs an analytical examination. Scopus has increased its coverage significantly in recent years and has paid special attention to journals from developed/emerging economies. Thus, Scopus is much more diversified now in terms of indexing journals from different countries. Therefore, one may be tempted to postulate that this is the underlying factor that has led to increasing positive association between journals indexed from a country and research output. However, a closer examination of the association between ratio of output in home journals (TPCJ) and total research output (TP) for various countries (Table 4) provide evidence that this postulate is not tenable. Table 4 indicates that many countries show a decline in proportion of papers published in their home journals during 2005 to 2019. Thus, while the number of journals for most of the countries has increased during this period, the proportion of papers published in home journals has not increased, or in other words, research output has grown at a rate faster than rate of growth of home journals and got expanded to include other internationally oriented journals too. One good example is China which published 63% of its research output in home journals in 2005, which reduced to only 17% in 2019. In the same period, China's research output has grown significantly, which would not have been possible unless China would have expanded its publication base and improved the quality of its research output to the standards of well-known international journals. India and Brazil are found to have largely maintained their research output proportion in home journals from 2005 to 2019, but at the same time have also grown significantly in terms of their total research output. Thus, it appears that other factors like internationalization of scientific publishing and growth in publication base (including growth in number of researchers) of different countries may also be playing an important role in growth of their research output.

Some previous studies have explored the reasons behind the growth of research outputs and have identified various factors behind the growth. For example, impact of Gross Expenditure on R&D (GERD) and Full Time Equivalent (FTE) have been examined in some studies (Coccia, 2009; Meo et al., 2013; Lancho-Barrantes, Ceballos-Cancino, & Cantu-Ortiz, 2020), while the impact of collaboration network structure on research output of a country was another factor to be examined (Guan et al., 2016). Bhattacharya et al. (2015) have explored the different changes that have happened in scientific research per se, looking at the endogenous and exogenous factors behind publication growth. Leta (2011), Najari & Yousefvand (2013) and Moed, Markusova & Akoev (2018) have underscored that the growth of number of journals from a country cannot alone be sufficient enough to explain the growth of research output for the countries. However, this study has dived deep to bring out how the journals from different countries are indexed in Scopus and how the pattern of indexing changes. The relationship that is seen from this granular level provides a deeper insight than that was possible in earlier research, inspite of the limitation of not being able to draw the underlying cause of the relationship (common factor) or whether one variable is affecting the other or vice-versa.

In the broader context of the discussion proposed above, it would also be relevant to discuss some policy perspectives about availability of suitable publication venues in different countries. There are several points that can be looked into. First, the lack of appropriate publication venues in certain countries results in missed opportunities of higher research output growth of the country. Secondly, concentration of majority of the journals in certain selected countries, indirectly guide the research agenda and theme in other countries. Many times, researchers ignore working on domestic and locally relevant problems, as such research work is less likely to find a place in the outside international journals. In this process many times the national context and local relevance of the research work is lost. Given that research has a special significance for the local and national context in which it is done and the problems that it solves, there should be enough venues for publishing such research work. Therefore, availability of more publication venues in a country should be an important goal for the science policy. This is specially more relevant for developing countries, which are expanding in terms of GERD, FTE etc. but perhaps not enough in terms of publication venues.

Another aspect that is closely related to this whole discussion is the current trend of journals to charge Article Processing Charges (APC) for publishing research papers. It has been observed that these charges are rooted in the so called 'global north' economic context and are exorbitant and unaffordable for researchers in the 'global south'. One interesting example in this context is the 'Read and Publish Agreement' in The Netherlands signed between the Association of Dutch Universities and the Koninklijke Nederlandse Akademie Van Nettenschappen. This agreement allows all participating institutions to publish Open Access in more than 2000 Open Access Hybrid journals of Springer with APC fees covered under the agreement along with full access to all Springer subscription journals. Similar arrangements are observed in North economies between the publishers and academic bodies. However, there are almost no institutional support mechanisms available in the developing countries to bear the APC. Therefore, if there are suitable number of publication venues in a country, they are more likely to be situated in the context and one may expect that such barriers of high APC may not be there with them. Accordingly, it should be an important policy consideration for developing countries to support development of enough publication venues for their researchers.

**Conclusion**

The study analysed the relationship between number of journals indexed from a country and its research output. Analytical results confirm a positive correlation between the number of journals (NCJ) and the research output (TP) of majority of the countries. Further, positive correlations are also observed in the growth rate of journals ($CAGR_J$) and papers ($CAGR_P$), and also in the growth of global share of number of journals (X1) and papers (X2) for a majority of the countries. There are, however, several countries which produce higher research output as compared to number of journals published from them and indexed in database. China, India, Brazil and Russia are some examples to mention. The proportionate share of publications in home journals has, however, declined for a good number of countries, indicating that other factors like wider publication base and internationalization of scientific publishing from those countries may also have a role to play.

Inspite of publishing becoming more concentrated globally with a few publishers dominating the global publication landscape; a country's affiliation of a journal is an important assertion of its scientific capacity. Journals develop over a long period to attract global attention. An influential journal has various quality attributes that lead towards making a global impact. *A journal is an institution in itself.* A reflected glory comes to a research organization/professional society/university if its journal is indexed in a global database and attracts scholarly research that provides new pathways for scientific research and creativity. Thus, there has to be dedicated support for a country to develop journals. The study suggests that the developing countries should focus on developing more publication venues for its researchers, both for providing suitable venues for publication of research output and also for supporting publication of research on domestic and locally relevant problems and issues. Development of such publication venues may also help in rationalizing the APCs, and perhaps improve the whole scientific publishing model.

**Limitations**

This study has analysed the relationship between number of journals published from different countries and their research output. In this process, we have taken the information of country of journals from the Scopus master journal list. It may be noted that some of the major commercial publishers- Elsevier, Nature, Springer etc.- have a large number of journals marked as being published from some specific countries, mainly US, UK, Germany and Netherlands. Therefore, the analytical results must be seen in this light and that there may be distortions in the relationships observed for these countries. Further, we have taken the publisher country as country of a journal. However, there are debates on what should be a good way to classify country of a journal, whether this should be the publisher country or the country which its Chief Editor (or the main editors) belong to. This work used the former and hence analytical results may be used with this understanding.

The second major limitation of this work is that it only analyses the relatedness of number of journals and research output for various countries and does not go deeper to analyse the pathways necessary to understand the attribution. It is quite clear that the growth of number of journals from a country cannot alone be sufficient enough to explain the growth of research output for all the countries. Such an attribution would need to analyse the pathways of the relationship between the two, in the notion of 'necessary' and 'sufficient' conditions of logical formalism. Existence of correlations cannot be taken as a 'necessary' and 'sufficient' conditions to attribute a causal relationship, and the results may therefore be seen in this light.

The third limitation is that this work only analyses the relationship between number of journals from a country and its research output and does not take into account other factors that are likely to affect the research output volume of the country. These factors may include GERD, FTE, collaboration networks etc. Different countries vary significantly on these aspects and hence merely looking at relationship between number of journals and research output of the country does not present the full picture. For example, China has grown from 850 researchers per million in 2005 to 1,350 researchers per million in 2018, which is a significant increase and would definitely be an important factor behind the phenomenal growth in China's research output. Similarly, the research funding volume of different countries also vary a lot, with US, UK, Germany spending higher proportion of their GDP on R&D activities, whereas many developing countries, such as India, continue to spend less than 1% of its GDP on R&D activities. The collaboration networks of different countries also vary a lot. Therefore, a detailed analysis of all these factors together, possibly as a multivariate regression model, can be taken up as a future work to understand the complex interplay of multiple factors determining research output of a country.

The study inspite of these limitations, has shown the changing profile of journals indexed from different countries; has examined in a granular level which enriches the understanding of the how journals are indexed in the Scopus database. It also refutes the simplistic interpretation of indicator of journal packing density that argued a causal factor of journal indexing in a home country to increase in research output. The study provides a more critical introspection of the relationship that is observed in a journal indexed from a country and its research output. In the process, it has identified many salient aspects of different countries research outputs and journals indexed.

**Conflict of Interest/ Ethical Declaration**

The authors have no conflicts of interest to declare that are relevant to the content of this article and the manuscript has full compliance with Ethical Standards indicated.

**Acknowledgements**

The authors would like to acknowledge that a pre-print version of this article appears in arXiv at: https://arxiv.org/abs/2103.11100 (Singh et al., 2021b).

**References**

Basu, A. (2010). Does a country's scientific 'productivity' depend critically on the number of country journals indexed? *Scientometrics*, *82*(3), 507-516.

Bhattacharya, S., Shilpa, Kaul, A. (2015). Emerging Countries Assertion in the Global Publication Landscape of Science: A Case Study of India. *Scientometrics*, 103(2): 387-411. http://link.springer.com/article/10.1007%2Fs11192-015-1551-4

Coccia, M. (2009). What is the optimal rate of R&D investment to maximize productivity growth?. *Technological Forecasting and Social Change*, 76(3), 433-446.

Collazo-Reyes, F. (2014). Growth of the number of indexed journals of Latin America and the Caribbean: the effect on the impact of each country. *Scientometrics*, *98*(1), 197-209.

Erfanmanesh, M., Tahira, M., & Abrizah, A. (2017). The publication success of 102 nations in Scopus and the performance of their Scopus-indexed journals. *Publishing Research Quarterly*, *33*(4), 421-432.

Gingras, Y., & Khelfaoui, M. (2018). Assessing the effect of the United States' "citation advantage" on other countries' scientific impact as measured in the Web of Science (WoS) database. *Scientometrics*, *114*(2), 517-532.

Guan, J., Zuo, K., Chen, K., & Yam, R. C. (2016). Does country-level R&D efficiency benefit from the collaboration network structure?. *Research Policy*, 45(4), 770-784.

Jinha, A.E. (2010). Article 50 million: an estimate of the number of scholarly articles in existence, *Learned Publishing*, 23:258–263 doi:10.1087/20100308

Lancho-Barrantes, B.S., Ceballos-Cancino, H.G. & Cantu-Ortiz, F.J. (2020) Comparing the efficiency of countries to assimilate and apply research investment. *Qual Quant*. https://doi.org/10.1007/s11135-020-01063-w

Leta, J. (2011). Growth of Brazilian science: A real internationalization or a matter of databases' coverage? In *Proceedings of International Conference of the International Society for Scientometrics and Informetrics*, South Africa, (392–408) July 2–7, 2011.

Meo, S. A., Al Masri, A. A., Usmani, A. M., Memon, A. N., & Zaidi, S. Z. (2013). Impact of GDP, spending on R&D, number of universities and scientific journals on research publications among Asian countries. *PloS one*, *8*(6), e66449.

Merton, R.K. (1963). Resistance to the systematic study of multiple discoveries in science. European *Journal of Sociology*, 4(2), 237-282

Michels, C., & Schmoch, U. (2012). The growth of science and database coverage. *Scientometrics*, 93(3), 831-846.

Moed, H. F., Markusova, V., & Akoev, M. (2018). Trends in Russian research output indexed in Scopus and Web of Science. *Scientometrics*, 116(2): 1153-1180.

Mongeon, P., & Paul-Hus, A. (2016). The journal coverage of Web of Science and Scopus: a comparative analysis. *Scientometrics*, *106*(1): 213-228.

Mueller, C. E. (2016). Accurate forecast of countries' research output by macro-level indicators. *Scientometrics*, *109*(2), 1307-1328.

Najari, A., & Yousefvand, M. (2013). Scientometrics study of impact of journal indexing on the growth of scientific productions of Iran. *Iranian Journal of public health*, 42(10): 1134.

Price, D. J. D. S. (1961). *Science Since Babylon*, Yale University Press, USA.

Singh P., Singh V.K., Arora P., Bhattacharya S. (2020). India's rank and global share in scientific research: how data drawn from different databases can produce varying outcomes? *Journal of Scientific and Industrial Research*, Vol. 80, No. 04, pp. 336-346.

Singh, V.K., Singh, P., Karmakar, M., Leta, J., Mayr, P. (2021a). The Journal Coverage of Web of Science, Scopus and Dimensions: A Comparative Analysis. *Scientometrics*, 126(6): 5113–5142. https://doi.org/10.1007/s11192-021-03948-5

Singh, V.K., Singh, P., Uddin, A., Arora, P. & Bhattacharya, S. (2021b) Influence of journals indexed from a country on its research output: An empirical investigation. Preprint available at: https://arxiv.org/abs/2103.11100

Tahamtan, I., Afshar, A. S., & Ahamdzadeh, K. (2016). Factors affecting number of citations: a comprehensive review of the literature. *Scientometrics*, *107*(3), 1195-1225.

# TABLES

## Table 1- year-wise number of journals (NCJ) for 50 countries and CAGR_J (2005-19)

| | NCJ=Number of Country Journal (active and inactive) | | | | | | | | | | | | | | | CAGR_J |
|---|---|---|---|---|---|---|---|---|---|---|---|---|---|---|---|---|
| | 2005 | 2006 | 2007 | 2008 | 2009 | 2010 | 2011 | 2012 | 2013 | 2014 | 2015 | 2016 | 2017 | 2018 | 2019 | |
| US | 4845 | 4967 | 5012 | 5130 | 5356 | 5538 | 5654 | 5681 | 5735 | 5832 | 5879 | 5969 | 5859 | 5854 | 5947 | **1.38** |
| China | 417 | 446 | 466 | 480 | 494 | 513 | 514 | 519 | 548 | 549 | 550 | 583 | 562 | 562 | 571 | **2.12** |
| UK | 3827 | 3987 | 4091 | 4266 | 4479 | 4718 | 4879 | 5001 | 5085 | 5216 | 5344 | 5418 | 5446 | 5551 | 5658 | **2.64** |
| Germany | 1094 | 1170 | 1209 | 1264 | 1357 | 1411 | 1485 | 1493 | 1507 | 1544 | 1558 | 1597 | 1586 | 1573 | 1592 | **2.53** |
| Japan | 376 | 382 | 390 | 401 | 411 | 414 | 425 | 431 | 427 | 415 | 419 | 414 | 383 | 371 | 374 | **-0.04** |
| India | 191 | 210 | 226 | 249 | 310 | 361 | 422 | 431 | 434 | 445 | 439 | 461 | 422 | 425 | 432 | **5.59** |
| France | 418 | 428 | 437 | 444 | 462 | 476 | 503 | 498 | 493 | 490 | 490 | 491 | 484 | 467 | 459 | **0.63** |
| Italy | 257 | 262 | 272 | 280 | 299 | 324 | 352 | 368 | 381 | 399 | 422 | 436 | 452 | 459 | 487 | **4.35** |
| Canada | 213 | 217 | 221 | 222 | 237 | 229 | 233 | 234 | 241 | 239 | 239 | 233 | 221 | 226 | 239 | **0.77** |
| Spain | 205 | 222 | 238 | 269 | 307 | 322 | 388 | 427 | 447 | 456 | 487 | 495 | 527 | 550 | 592 | **7.33** |
| Australia | 102 | 108 | 111 | 138 | 164 | 171 | 185 | 180 | 175 | 173 | 172 | 183 | 174 | 166 | 174 | **3.62** |
| South Korea | 51 | 61 | 83 | 110 | 124 | 147 | 169 | 182 | 186 | 193 | 203 | 212 | 227 | 247 | 264 | **11.58** |
| Brazil | 95 | 157 | 188 | 219 | 235 | 258 | 282 | 303 | 311 | 318 | 327 | 336 | 347 | 362 | 375 | **9.59** |
| Russia | 240 | 190 | 196 | 210 | 239 | 244 | 256 | 267 | 290 | 302 | 325 | 386 | 418 | 451 | 506 | **5.10** |
| Netherlands | 1553 | 1571 | 1616 | 1671 | 1705 | 1733 | 1821 | 1870 | 1927 | 2009 | 2051 | 2066 | 2052 | 2038 | 2064 | **1.91** |
| Iran | 21 | 26 | 30 | 43 | 59 | 78 | 101 | 103 | 110 | 117 | 129 | 142 | 148 | 157 | 178 | **15.31** |
| Turkey | 79 | 85 | 88 | 112 | 138 | 152 | 165 | 162 | 169 | 174 | 182 | 192 | 183 | 186 | 197 | **6.28** |
| Switzerland | 282 | 293 | 324 | 359 | 390 | 430 | 470 | 492 | 513 | 548 | 605 | 640 | 671 | 678 | 695 | **6.20** |
| Poland | 174 | 187 | 198 | 222 | 249 | 252 | 268 | 279 | 288 | 292 | 299 | 318 | 321 | 343 | 371 | **5.18** |
| Taiwan | 44 | 45 | 50 | 52 | 63 | 63 | 69 | 73 | 70 | 69 | 73 | 75 | 79 | 83 | 86 | **4.57** |
| Sweden | 24 | 26 | 27 | 32 | 34 | 33 | 39 | 41 | 41 | 43 | 45 | 43 | 42 | 41 | 45 | **4.28** |
| Belgium | 94 | 96 | 95 | 100 | 102 | 104 | 125 | 123 | 117 | 122 | 117 | 114 | 117 | 112 | 117 | **1.47** |
| Denmark | 19 | 18 | 17 | 20 | 21 | 22 | 29 | 28 | 25 | 33 | 31 | 30 | 27 | 29 | 30 | **3.09** |
| Austria | 28 | 33 | 36 | 35 | 38 | 41 | 49 | 48 | 48 | 48 | 49 | 51 | 47 | 44 | 46 | **3.36** |
| Malaysia | 11 | 17 | 27 | 36 | 46 | 53 | 67 | 72 | 73 | 75 | 77 | 81 | 92 | 96 | 99 | **15.78** |
| Czech Republic | 88 | 91 | 100 | 104 | 115 | 121 | 138 | 142 | 152 | 160 | 166 | 176 | 180 | 179 | 186 | **5.12** |

| Country | | | | | | | | | | | | | | | CAGR |
|---|---|---|---|---|---|---|---|---|---|---|---|---|---|---|---|
| Israel | 14 | 12 | 12 | 13 | 13 | 13 | 12 | 12 | 12 | 12 | 12 | 12 | 9 | 8 | 9 | **-2.90** |
| Portugal | 10 | 11 | 14 | 17 | 21 | 21 | 31 | 32 | 38 | 39 | 42 | 47 | 49 | 54 | 63 | **13.05** |
| Mexico | 41 | 44 | 47 | 69 | 73 | 73 | 81 | 89 | 93 | 102 | 101 | 103 | 105 | 106 | 107 | **6.60** |
| Norway | 14 | 13 | 13 | 14 | 15 | 14 | 22 | 20 | 22 | 25 | 24 | 25 | 25 | 27 | 29 | **4.97** |
| Greece | 23 | 25 | 25 | 36 | 42 | 48 | 53 | 56 | 54 | 55 | 50 | 47 | 51 | 49 | 47 | **4.88** |
| Finland | 29 | 33 | 31 | 30 | 33 | 32 | 43 | 42 | 42 | 44 | 39 | 39 | 40 | 35 | 36 | **1.45** |
| Hong Kong | 17 | 17 | 16 | 16 | 20 | 19 | 20 | 23 | 21 | 22 | 22 | 22 | 22 | 21 | 21 | **1.42** |
| Singapore | 78 | 84 | 87 | 98 | 104 | 108 | 105 | 108 | 106 | 108 | 111 | 116 | 125 | 129 | 134 | **3.67** |
| South Africa | 38 | 38 | 41 | 53 | 56 | 59 | 63 | 64 | 66 | 67 | 69 | 72 | 71 | 73 | 81 | **5.18** |
| New Zealand | 24 | 26 | 25 | 39 | 50 | 59 | 60 | 62 | 58 | 62 | 66 | 66 | 61 | 56 | 57 | **5.94** |
| Egypt | 44 | 49 | 63 | 75 | 105 | 153 | 164 | 179 | 185 | 202 | 209 | 217 | 204 | 189 | 194 | **10.40** |
| Saudi Arabia | 5 | 6 | 6 | 7 | 19 | 19 | 21 | 20 | 20 | 19 | 19 | 19 | 19 | 19 | 20 | **9.68** |
| Romania | 27 | 28 | 33 | 59 | 86 | 100 | 125 | 137 | 152 | 155 | 166 | 172 | 171 | 175 | 178 | **13.40** |
| Ireland | 11 | 12 | 12 | 13 | 13 | 15 | 18 | 16 | 21 | 20 | 20 | 20 | 22 | 20 | 22 | **4.73** |
| Thailand | 8 | 12 | 12 | 18 | 22 | 22 | 27 | 28 | 28 | 27 | 29 | 28 | 39 | 46 | 57 | **13.99** |
| Argentina | 23 | 26 | 28 | 40 | 42 | 45 | 49 | 53 | 52 | 55 | 56 | 58 | 57 | 59 | 64 | **7.06** |
| Pakistan | 15 | 32 | 34 | 44 | 71 | 80 | 92 | 97 | 94 | 94 | 89 | 88 | 73 | 75 | 54 | **8.91** |
| Indonesia | 3 | 3 | 3 | 3 | 6 | 10 | 14 | 18 | 19 | 25 | 29 | 33 | 42 | 52 | 57 | **21.69** |
| Hungary | 55 | 56 | 78 | 86 | 87 | 86 | 97 | 97 | 95 | 93 | 92 | 97 | 93 | 94 | 93 | **3.56** |
| Ukraine | 27 | 16 | 17 | 23 | 27 | 28 | 30 | 35 | 39 | 38 | 44 | 44 | 41 | 46 | 57 | **5.11** |
| Chile | 19 | 35 | 49 | 66 | 67 | 68 | 72 | 78 | 81 | 84 | 87 | 93 | 96 | 98 | 103 | **11.93** |
| Colombia | 12 | 13 | 22 | 37 | 46 | 49 | 57 | 68 | 79 | 85 | 88 | 94 | 97 | 108 | 114 | **16.19** |
| Slovakia | 29 | 26 | 29 | 31 | 37 | 38 | 43 | 44 | 45 | 45 | 46 | 50 | 56 | 59 | 65 | **5.53** |
| Serbia | 12 | 13 | 28 | 33 | 37 | 41 | 45 | 51 | 51 | 53 | 55 | 59 | 63 | 67 | 74 | **12.89** |
| **World** | **15606** | **16250** | **16832** | **17842** | **19069** | **19980** | **21076** | **21525** | **21924** | **22450** | **22917** | **23455** | **23409** | **23640** | **24278** | **2.99** |

Note: NCJ- Number of journals from a country; CAGR- Compounded Annual Growth Rate

**Table 2: year-wise number of total papers (TP) of 50 countries and their CAGR$_P$ (2005-19)**

| | TP=TPCJ+TPOJ | | | | | | | | | | | | | | | CAGR$_P$ |
|---|---|---|---|---|---|---|---|---|---|---|---|---|---|---|---|---|
| | 2005 | 2006 | 2007 | 2008 | 2009 | 2010 | 2011 | 2012 | 2013 | 2014 | 2015 | 2016 | 2017 | 2018 | 2019 | |
| US | 348248 | 366272 | 371282 | 380164 | 397417 | 407883 | 426866 | 445537 | 460780 | 471311 | 475637 | 481551 | 490437 | 502160 | 516612 | **2.66** |
| China | 141750 | 165402 | 178564 | 198408 | 220183 | 230892 | 253777 | 286180 | 332272 | 366824 | 388257 | 416343 | 444463 | 495549 | 564672 | **9.65** |
| UK | 95164 | 103814 | 108306 | 108802 | 114057 | 116550 | 122055 | 128567 | 136372 | 138709 | 142996 | 147372 | 150662 | 157059 | 165463 | **3.76** |
| Germany | 87137 | 93059 | 94884 | 97128 | 101764 | 104861 | 110466 | 116583 | 120813 | 124724 | 125999 | 130153 | 133139 | 135224 | 138114 | **3.12** |
| Japan | 87307 | 91469 | 90180 | 91507 | 93900 | 90475 | 93229 | 95648 | 97981 | 95950 | 93931 | 96747 | 97085 | 98617 | 100373 | **0.93** |
| India | 33428 | 38410 | 42763 | 47863 | 54627 | 63009 | 74893 | 81519 | 88358 | 100049 | 106566 | 110942 | 109663 | 118503 | 142083 | **10.13** |
| France | 58205 | 63298 | 65580 | 69725 | 74058 | 75391 | 78707 | 82040 | 85857 | 87866 | 87945 | 91325 | 91718 | 92005 | 91358 | **3.05** |
| Italy | 45123 | 49383 | 52939 | 55940 | 60068 | 60572 | 64412 | 69711 | 75879 | 78981 | 80775 | 84096 | 86147 | 90451 | 94936 | **5.08** |
| Canada | 49301 | 53584 | 56050 | 58797 | 62838 | 64329 | 67162 | 71140 | 74598 | 77120 | 78350 | 80003 | 82340 | 86146 | 91145 | **4.18** |
| Spain | 36092 | 41349 | 44139 | 47683 | 51634 | 54126 | 59494 | 64057 | 67742 | 70269 | 70518 | 73538 | 75535 | 78225 | 84238 | **5.81** |
| Australia | 32724 | 36724 | 38718 | 42186 | 46388 | 49802 | 53660 | 58151 | 65220 | 69873 | 73304 | 76037 | 78285 | 82956 | 89357 | **6.93** |
| South Korea | 24596 | 28056 | 31658 | 36453 | 40844 | 44911 | 48905 | 54574 | 59504 | 64649 | 67805 | 69407 | 70414 | 72294 | 76472 | **7.86** |
| Brazil | 20630 | 27200 | 30767 | 35253 | 38706 | 40864 | 44670 | 49159 | 51910 | 55402 | 57237 | 61571 | 65015 | 69347 | 72872 | **8.78** |
| Russia | 31127 | 28068 | 29727 | 31836 | 32975 | 32813 | 35832 | 35325 | 40368 | 45706 | 53833 | 61452 | 65227 | 71850 | 77561 | **6.28** |
| Netherlands | 26538 | 28524 | 29764 | 30964 | 34247 | 36182 | 37990 | 41309 | 43525 | 44847 | 45185 | 46655 | 47443 | 49258 | 51701 | **4.55** |
| Iran | 6202 | 8672 | 11752 | 14988 | 19171 | 22667 | 31597 | 34677 | 37244 | 40501 | 41103 | 47581 | 50188 | 53606 | 59474 | **16.27** |
| Turkey | 18013 | 19819 | 21545 | 22754 | 26074 | 27194 | 28971 | 30477 | 32951 | 33862 | 35706 | 38167 | 35293 | 36493 | 41570 | **5.73** |
| Switzerland | 18952 | 20874 | 21862 | 22713 | 24369 | 25572 | 27809 | 30273 | 32031 | 33580 | 34289 | 35841 | 37419 | 38114 | 39132 | **4.95** |
| Poland | 18575 | 20927 | 21131 | 22883 | 24185 | 24529 | 26333 | 29000 | 31392 | 32720 | 34516 | 36387 | 36271 | 37534 | 39907 | **5.23** |
| Taiwan | 17656 | 20688 | 22082 | 24620 | 26646 | 27499 | 29540 | 30858 | 31541 | 30730 | 29109 | 28627 | 27382 | 27246 | 29266 | **3.43** |
| Sweden | 18433 | 19619 | 20069 | 20247 | 22023 | 22707 | 24054 | 25993 | 27928 | 29766 | 30992 | 32182 | 33228 | 34355 | 35714 | **4.51** |
| Belgium | 14456 | 15230 | 16348 | 17466 | 18898 | 19544 | 20935 | 22423 | 23918 | 25319 | 25694 | 26268 | 26831 | 27709 | 27682 | **4.43** |
| Denmark | 9891 | 10638 | 11087 | 11771 | 12786 | 13806 | 15318 | 16740 | 17986 | 20142 | 20809 | 21669 | 22390 | 23279 | 24084 | **6.11** |
| Austria | 9854 | 10616 | 11381 | 12007 | 12835 | 13405 | 14668 | 15447 | 16543 | 17597 | 18181 | 18874 | 19312 | 20160 | 20984 | **5.17** |
| Malaysia | 2374 | 3028 | 3541 | 4897 | 7698 | 9862 | 13295 | 14828 | 17095 | 18844 | 21100 | 22325 | 22799 | 24023 | 27535 | **17.75** |
| Czech Republic | 7404 | 8736 | 9607 | 10008 | 10795 | 11669 | 12857 | 13757 | 14606 | 15951 | 17131 | 17739 | 18275 | 18633 | 19502 | **6.67** |
| Israel | 11428 | 12482 | 12608 | 12889 | 13086 | 13120 | 13620 | 14293 | 14660 | 15433 | 15763 | 16543 | 16715 | 17572 | 18195 | **3.15** |

| Country | | | | | | | | | | | | | | | | |
|---|---|---|---|---|---|---|---|---|---|---|---|---|---|---|---|---|
| Portugal | 5693 | 7106 | 7491 | 8799 | 9866 | 10789 | 12340 | 13964 | 15609 | 16692 | 17126 | 18086 | 18575 | 19585 | 21356 | **9.21** |
| Mexico | 8247 | 9003 | 9657 | 10761 | 11679 | 11708 | 12785 | 14222 | 15137 | 16481 | 17174 | 18353 | 19846 | 21380 | 23232 | **7.15** |
| Norway | 7571 | 8417 | 9011 | 9586 | 10879 | 11241 | 12410 | 13310 | 13812 | 14718 | 15008 | 16294 | 17110 | 18472 | 19524 | **6.52** |
| Greece | 9051 | 10530 | 11248 | 11815 | 12912 | 12621 | 13032 | 13211 | 13432 | 13604 | 13117 | 13692 | 13702 | 14417 | 15115 | **3.48** |
| Finland | 9186 | 9880 | 10097 | 10464 | 11131 | 11415 | 12180 | 12850 | 13551 | 14662 | 15127 | 15473 | 15730 | 16181 | 17142 | **4.25** |
| Hong Kong | 8746 | 9643 | 9767 | 9794 | 10615 | 10538 | 11320 | 12034 | 13278 | 14077 | 14326 | 15157 | 16191 | 17904 | 19149 | **5.36** |
| Singapore | 6827 | 7534 | 7607 | 8292 | 9124 | 9949 | 10739 | 11948 | 13086 | 13942 | 14708 | 15170 | 15875 | 16364 | 17237 | **6.37** |
| South Africa | 6204 | 7146 | 7242 | 8073 | 9062 | 9863 | 11327 | 12459 | 13844 | 16620 | 16370 | 18043 | 19174 | 20483 | 22664 | **9.02** |
| New Zealand | 6790 | 7162 | 7642 | 7858 | 8538 | 9245 | 10219 | 10470 | 11016 | 11543 | 11731 | 12092 | 12671 | 13140 | 13877 | **4.88** |
| Egypt | 3951 | 4351 | 4827 | 5481 | 6949 | 7576 | 9509 | 11076 | 12413 | 13339 | 14378 | 16366 | 15894 | 18579 | 21452 | **11.94** |
| Saudi Arabia | 1955 | 2162 | 2261 | 2579 | 3496 | 4970 | 7603 | 9864 | 12263 | 14956 | 16431 | 17813 | 18094 | 19679 | 23439 | **18.01** |
| Romania | 3132 | 3463 | 4083 | 6183 | 7871 | 8295 | 9340 | 9812 | 10814 | 10198 | 10750 | 10497 | 10077 | 10205 | 11363 | **8.97** |
| Ireland | 4702 | 5467 | 5945 | 6536 | 7441 | 8234 | 8832 | 8916 | 9311 | 9755 | 9649 | 10334 | 10748 | 11687 | 12490 | **6.73** |
| Thailand | 3737 | 4667 | 5131 | 5730 | 6484 | 7063 | 7850 | 8806 | 9071 | 9715 | 10175 | 11365 | 11959 | 13883 | 14957 | **9.69** |
| Argentina | 5878 | 6538 | 7047 | 7684 | 8640 | 9120 | 9731 | 10198 | 10601 | 11183 | 11348 | 11699 | 12103 | 13042 | 13009 | **5.44** |
| Pakistan | 2233 | 2929 | 3479 | 4205 | 5184 | 6069 | 7915 | 8399 | 9755 | 10180 | 10667 | 12464 | 14858 | 17194 | 20333 | **15.87** |
| Indonesia | 885 | 996 | 1006 | 1103 | 1480 | 1815 | 2283 | 2685 | 3284 | 4257 | 5850 | 7826 | 10464 | 13979 | 20859 | **23.45** |
| Hungary | 5727 | 6065 | 6403 | 6756 | 7048 | 6813 | 7683 | 8038 | 8033 | 8604 | 8695 | 8917 | 9108 | 9614 | 9793 | **3.64** |
| Ukraine | 5563 | 4968 | 4969 | 5413 | 5551 | 5468 | 6248 | 6809 | 7557 | 8595 | 8546 | 8752 | 9337 | 10536 | 11652 | **5.05** |
| Chile | 3368 | 3980 | 4456 | 4921 | 5598 | 5853 | 6632 | 7442 | 7946 | 9276 | 10034 | 11124 | 11606 | 12854 | 13793 | **9.85** |
| Colombia | 1272 | 1715 | 1987 | 2874 | 3474 | 3901 | 4370 | 5187 | 5693 | 6313 | 7077 | 8243 | 9242 | 10379 | 11270 | **15.65** |
| Slovakia | 2606 | 2999 | 3181 | 3493 | 3517 | 3741 | 4129 | 4349 | 4843 | 5341 | 5436 | 6104 | 6071 | 6082 | 6489 | **6.27** |
| Serbia | 1452 | 1607 | 1777 | 3304 | 3882 | 4403 | 5062 | 6365 | 6294 | 6251 | 6189 | 6428 | 6531 | 6688 | 7049 | **11.11** |
| **World** | **1291181** | **1396113** | **1464044** | **1513234** | **1602123** | **1646399** | **1761770** | **1848900** | **1968219** | **2055870** | **2092607** | **2154451** | **2240691** | **2293203** | **2472338** | **4.43** |

**Note:** TP- Total papers; TPCJ- Total papers in home journals; TPOJ- Total papers in outside journals.

**Table 3: Pearson Correlation Coefficients of NCJ vs TP for all the 50 countries**

| Country | Pearson Correlation Coefficient (TP vs NCJ) for 15 years |
|---|---|
| US | 0.96 |
| China | 0.90 |
| UK | 0.97 |
| Germany | 0.95 |
| Japan | 0.03 |
| India | 0.87 |
| France | 0.76 |
| Italy | 0.99 |
| Canada | 0.61 |
| Spain | 1.00 |
| Australia | 0.73 |
| South Korea | 0.98 |
| Brazil | 0.97 |
| Russia | 0.99 |
| Netherlands | 0.98 |
| Iran | 1.00 |
| Turkey | 0.97 |
| Switzerland | 0.99 |
| Poland | 0.97 |
| Taiwan | 0.80 |
| Sweden | 0.90 |
| Belgium | 0.79 |
| Denmark | 0.87 |
| Austria | 0.82 |
| Malaysia | 0.97 |
| Czech Republic | 1.00 |
| Israel | -0.84 |
| Portugal | 0.99 |
| Mexico | 0.92 |
| Norway | 0.96 |
| Greece | 0.83 |
| Finland | 0.59 |
| Hong Kong | 0.70 |
| Singapore | 0.93 |
| South Africa | 0.93 |
| New Zealand | 0.81 |
| Egypt | 0.87 |
| Saudi Arabia | 0.68 |
| Romania | 0.97 |
| Ireland | 0.93 |
| Thailand | 0.96 |
| Argentina | 0.97 |
| Pakistan | 0.43 |
| Indonesia | 0.95 |
| Hungary | 0.77 |
| Ukraine | 0.95 |
| Chile | 0.91 |
| Colombia | 0.97 |
| Slovakia | 0.95 |
| Serbia | 0.96 |

**Average value = 0.84**

**Note:** NCJ- Number of journals from a country; TP- Total papers.

**Table 4: year-wise Ratio of TPCJ by TP for all the 50 countries**

| | TPCJ/TP | | | | | | | | | | | | | | |
|---|---|---|---|---|---|---|---|---|---|---|---|---|---|---|---|
| | 2005 | 2006 | 2007 | 2008 | 2009 | 2010 | 2011 | 2012 | 2013 | 2014 | 2015 | 2016 | 2017 | 2018 | 2019 |
| US | 0.56 | 0.55 | 0.53 | 0.52 | 0.53 | 0.53 | 0.52 | 0.53 | 0.52 | 0.50 | 0.50 | 0.49 | 0.48 | 0.47 | 0.46 |
| China | 0.63 | 0.58 | 0.55 | 0.52 | 0.49 | 0.46 | 0.40 | 0.36 | 0.34 | 0.30 | 0.27 | 0.25 | 0.22 | 0.20 | 0.17 |
| UK | 0.44 | 0.43 | 0.42 | 0.41 | 0.42 | 0.42 | 0.42 | 0.41 | 0.41 | 0.41 | 0.42 | 0.42 | 0.42 | 0.41 | 0.41 |
| Germany | 0.24 | 0.22 | 0.22 | 0.21 | 0.21 | 0.21 | 0.21 | 0.21 | 0.20 | 0.20 | 0.19 | 0.19 | 0.19 | 0.17 | 0.16 |
| Japan | 0.27 | 0.26 | 0.24 | 0.23 | 0.25 | 0.26 | 0.24 | 0.24 | 0.22 | 0.20 | 0.20 | 0.20 | 0.18 | 0.17 | 0.16 |
| India | 0.27 | 0.24 | 0.23 | 0.24 | 0.25 | 0.28 | 0.29 | 0.28 | 0.28 | 0.28 | 0.32 | 0.29 | 0.24 | 0.22 | 0.27 |
| France | 0.18 | 0.17 | 0.16 | 0.16 | 0.17 | 0.17 | 0.16 | 0.15 | 0.14 | 0.14 | 0.13 | 0.13 | 0.12 | 0.11 | 0.09 |
| Italy | 0.12 | 0.10 | 0.10 | 0.09 | 0.09 | 0.09 | 0.09 | 0.10 | 0.09 | 0.09 | 0.09 | 0.09 | 0.09 | 0.09 | 0.08 |
| Canada | 0.07 | 0.07 | 0.06 | 0.06 | 0.06 | 0.06 | 0.05 | 0.05 | 0.05 | 0.05 | 0.04 | 0.04 | 0.04 | 0.04 | 0.04 |
| Spain | 0.18 | 0.18 | 0.16 | 0.16 | 0.16 | 0.16 | 0.16 | 0.16 | 0.16 | 0.15 | 0.15 | 0.15 | 0.15 | 0.15 | 0.14 |
| Australia | 0.07 | 0.07 | 0.06 | 0.07 | 0.06 | 0.06 | 0.06 | 0.05 | 0.05 | 0.04 | 0.04 | 0.04 | 0.04 | 0.04 | 0.03 |
| South Korea | 0.15 | 0.15 | 0.18 | 0.18 | 0.19 | 0.21 | 0.20 | 0.19 | 0.19 | 0.18 | 0.18 | 0.17 | 0.16 | 0.17 | 0.15 |
| Brazil | 0.29 | 0.36 | 0.38 | 0.38 | 0.39 | 0.39 | 0.39 | 0.37 | 0.34 | 0.32 | 0.30 | 0.28 | 0.27 | 0.27 | 0.26 |
| Russia | 0.51 | 0.45 | 0.45 | 0.47 | 0.49 | 0.53 | 0.50 | 0.47 | 0.45 | 0.44 | 0.43 | 0.45 | 0.48 | 0.49 | 0.48 |
| Netherlands | 0.22 | 0.22 | 0.22 | 0.20 | 0.21 | 0.20 | 0.19 | 0.19 | 0.20 | 0.19 | 0.19 | 0.19 | 0.18 | 0.18 | 0.17 |
| Iran | 0.12 | 0.09 | 0.10 | 0.11 | 0.12 | 0.12 | 0.12 | 0.13 | 0.13 | 0.13 | 0.13 | 0.13 | 0.13 | 0.12 | 0.11 |
| Turkey | 0.15 | 0.15 | 0.14 | 0.16 | 0.19 | 0.22 | 0.24 | 0.24 | 0.21 | 0.20 | 0.20 | 0.19 | 0.19 | 0.16 | 0.15 |
| Switzerland | 0.08 | 0.07 | 0.06 | 0.06 | 0.06 | 0.06 | 0.06 | 0.07 | 0.07 | 0.07 | 0.07 | 0.07 | 0.08 | 0.08 | 0.10 |
| Poland | 0.39 | 0.38 | 0.36 | 0.37 | 0.38 | 0.38 | 0.37 | 0.37 | 0.33 | 0.29 | 0.27 | 0.27 | 0.26 | 0.23 | 0.20 |
| Taiwan | 0.08 | 0.07 | 0.07 | 0.06 | 0.07 | 0.07 | 0.07 | 0.07 | 0.06 | 0.06 | 0.06 | 0.06 | 0.06 | 0.06 | 0.05 |
| Sweden | 0.03 | 0.03 | 0.02 | 0.02 | 0.03 | 0.03 | 0.02 | 0.02 | 0.01 | 0.02 | 0.02 | 0.02 | 0.02 | 0.01 | 0.01 |
| Belgium | 0.07 | 0.06 | 0.06 | 0.06 | 0.05 | 0.05 | 0.05 | 0.05 | 0.04 | 0.04 | 0.03 | 0.03 | 0.03 | 0.02 | 0.02 |
| Denmark | 0.04 | 0.04 | 0.04 | 0.05 | 0.05 | 0.05 | 0.04 | 0.04 | 0.04 | 0.05 | 0.03 | 0.02 | 0.02 | 0.01 | 0.01 |
| Austria | 0.02 | 0.02 | 0.02 | 0.02 | 0.02 | 0.02 | 0.02 | 0.02 | 0.02 | 0.02 | 0.02 | 0.02 | 0.02 | 0.02 | 0.01 |
| Malaysia | 0.11 | 0.13 | 0.12 | 0.14 | 0.11 | 0.11 | 0.10 | 0.11 | 0.12 | 0.12 | 0.17 | 0.16 | 0.17 | 0.19 | 0.14 |
| Czech Republic | 0.33 | 0.30 | 0.30 | 0.27 | 0.26 | 0.27 | 0.28 | 0.27 | 0.24 | 0.22 | 0.22 | 0.20 | 0.19 | 0.18 | 0.17 |
| Israel | 0.03 | 0.04 | 0.03 | 0.03 | 0.03 | 0.02 | 0.02 | 0.02 | 0.02 | 0.01 | 0.01 | 0.02 | 0.01 | 0.01 | 0.01 |
| Portugal | 0.06 | 0.05 | 0.05 | 0.05 | 0.05 | 0.05 | 0.06 | 0.05 | 0.04 | 0.05 | 0.04 | 0.03 | 0.04 | 0.04 | 0.04 |
| Mexico | 0.15 | 0.16 | 0.15 | 0.17 | 0.17 | 0.17 | 0.18 | 0.17 | 0.16 | 0.17 | 0.15 | 0.13 | 0.12 | 0.12 | 0.11 |
| Norway | 0.05 | 0.04 | 0.04 | 0.03 | 0.04 | 0.03 | 0.03 | 0.03 | 0.03 | 0.02 | 0.02 | 0.02 | 0.02 | 0.03 | 0.03 |
| Greece | 0.06 | 0.06 | 0.04 | 0.05 | 0.05 | 0.05 | 0.05 | 0.05 | 0.04 | 0.04 | 0.04 | 0.05 | 0.04 | 0.04 | 0.04 |
| Finland | 0.05 | 0.03 | 0.02 | 0.02 | 0.04 | 0.03 | 0.03 | 0.03 | 0.02 | 0.02 | 0.01 | 0.01 | 0.01 | 0.01 | 0.01 |
| Hong Kong | 0.03 | 0.03 | 0.03 | 0.03 | 0.03 | 0.02 | 0.02 | 0.02 | 0.01 | 0.01 | 0.01 | 0.01 | 0.01 | 0.01 | 0.01 |
| Singapore | 0.05 | 0.04 | 0.04 | 0.03 | 0.04 | 0.03 | 0.03 | 0.03 | 0.03 | 0.02 | 0.02 | 0.02 | 0.02 | 0.02 | 0.02 |

| Country | | | | | | | | | | | | | | |
|---|---|---|---|---|---|---|---|---|---|---|---|---|---|---|
| **South Africa** | 0.15 | 0.15 | 0.13 | 0.13 | 0.13 | 0.12 | 0.13 | 0.12 | 0.12 | 0.11 | 0.12 | 0.12 | 0.11 | 0.10 | 0.11 |
| **New Zealand** | 0.05 | 0.05 | 0.05 | 0.06 | 0.06 | 0.05 | 0.06 | 0.04 | 0.05 | 0.04 | 0.04 | 0.04 | 0.04 | 0.03 | 0.04 |
| **Egypt** | 0.10 | 0.07 | 0.06 | 0.06 | 0.05 | 0.04 | 0.04 | 0.05 | 0.06 | 0.06 | 0.05 | 0.05 | 0.04 | 0.05 | 0.05 |
| **Saudi Arabia** | 0.14 | 0.11 | 0.09 | 0.10 | 0.09 | 0.08 | 0.05 | 0.04 | 0.03 | 0.03 | 0.03 | 0.02 | 0.03 | 0.03 | 0.03 |
| **Romania** | 0.37 | 0.33 | 0.34 | 0.46 | 0.50 | 0.49 | 0.47 | 0.42 | 0.42 | 0.39 | 0.40 | 0.39 | 0.36 | 0.33 | 0.32 |
| **Ireland** | 0.03 | 0.03 | 0.03 | 0.02 | 0.03 | 0.02 | 0.02 | 0.03 | 0.02 | 0.02 | 0.02 | 0.02 | 0.02 | 0.02 | 0.02 |
| **Thailand** | 0.23 | 0.21 | 0.18 | 0.16 | 0.16 | 0.15 | 0.15 | 0.16 | 0.12 | 0.13 | 0.12 | 0.11 | 0.11 | 0.14 | 0.12 |
| **Argentina** | 0.10 | 0.09 | 0.09 | 0.10 | 0.11 | 0.11 | 0.11 | 0.09 | 0.09 | 0.09 | 0.09 | 0.09 | 0.08 | 0.08 | 0.08 |
| **Pakistan** | 0.43 | 0.41 | 0.36 | 0.35 | 0.34 | 0.31 | 0.31 | 0.32 | 0.31 | 0.26 | 0.23 | 0.22 | 0.19 | 0.18 | 0.14 |
| **Indonesia** | 0.14 | 0.13 | 0.12 | 0.10 | 0.13 | 0.15 | 0.17 | 0.18 | 0.14 | 0.12 | 0.13 | 0.13 | 0.14 | 0.13 | 0.10 |
| **Hungary** | 0.22 | 0.21 | 0.23 | 0.23 | 0.22 | 0.22 | 0.23 | 0.21 | 0.18 | 0.19 | 0.17 | 0.16 | 0.15 | 0.16 | 0.13 |
| **Ukraine** | 0.16 | 0.06 | 0.06 | 0.10 | 0.12 | 0.14 | 0.11 | 0.19 | 0.22 | 0.24 | 0.29 | 0.27 | 0.26 | 0.25 | 0.24 |
| **Chile** | 0.15 | 0.21 | 0.21 | 0.24 | 0.25 | 0.24 | 0.23 | 0.22 | 0.21 | 0.19 | 0.17 | 0.16 | 0.14 | 0.13 | 0.12 |
| **Colombia** | 0.16 | 0.19 | 0.23 | 0.31 | 0.33 | 0.32 | 0.32 | 0.30 | 0.27 | 0.26 | 0.23 | 0.24 | 0.19 | 0.17 | 0.16 |
| **Slovakia** | 0.23 | 0.19 | 0.17 | 0.18 | 0.17 | 0.18 | 0.16 | 0.16 | 0.18 | 0.16 | 0.14 | 0.14 | 0.14 | 0.13 | 0.15 |
| **Serbia** | 0.29 | 0.23 | 0.36 | 0.27 | 0.26 | 0.25 | 0.20 | 0.19 | 0.20 | 0.19 | 0.18 | 0.18 | 0.18 | 0.16 | 0.16 |

**Note:** TPCJ- Total papers in home journals; TP- Total papers.

**FIGURES**

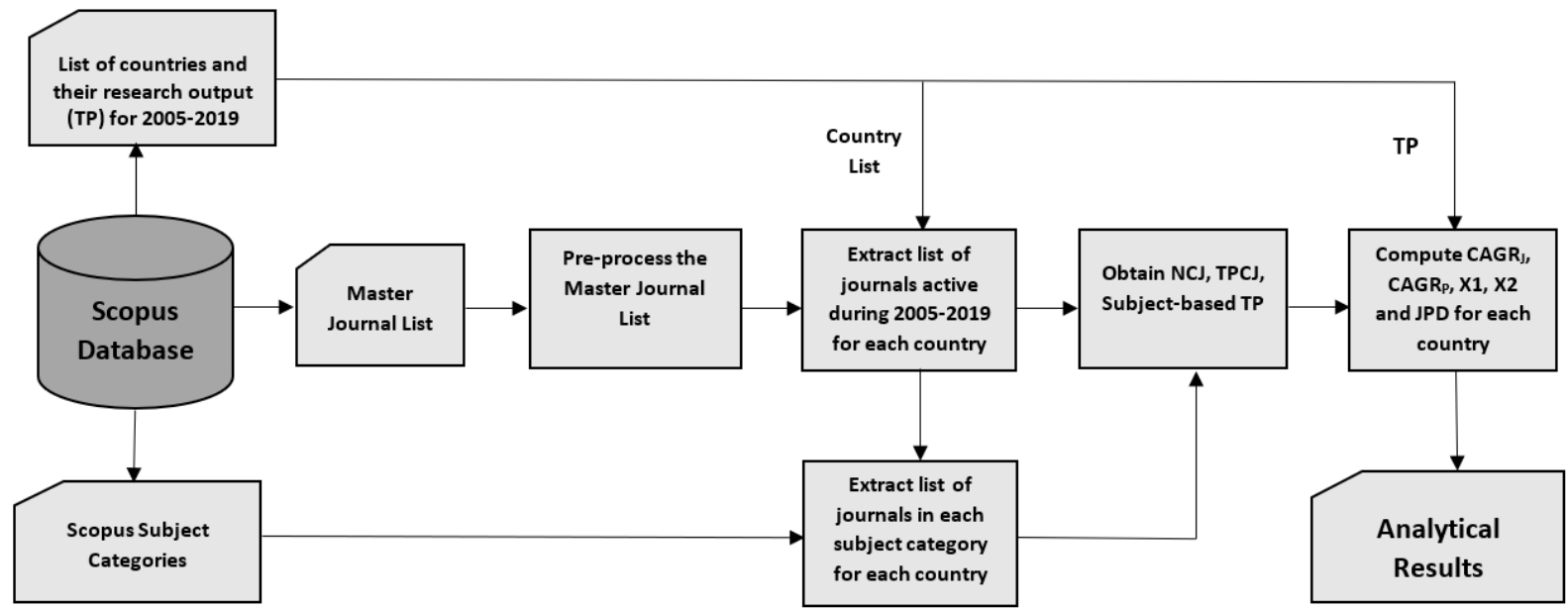

**Figure 1:** Schematic diagram of the analytical methodology

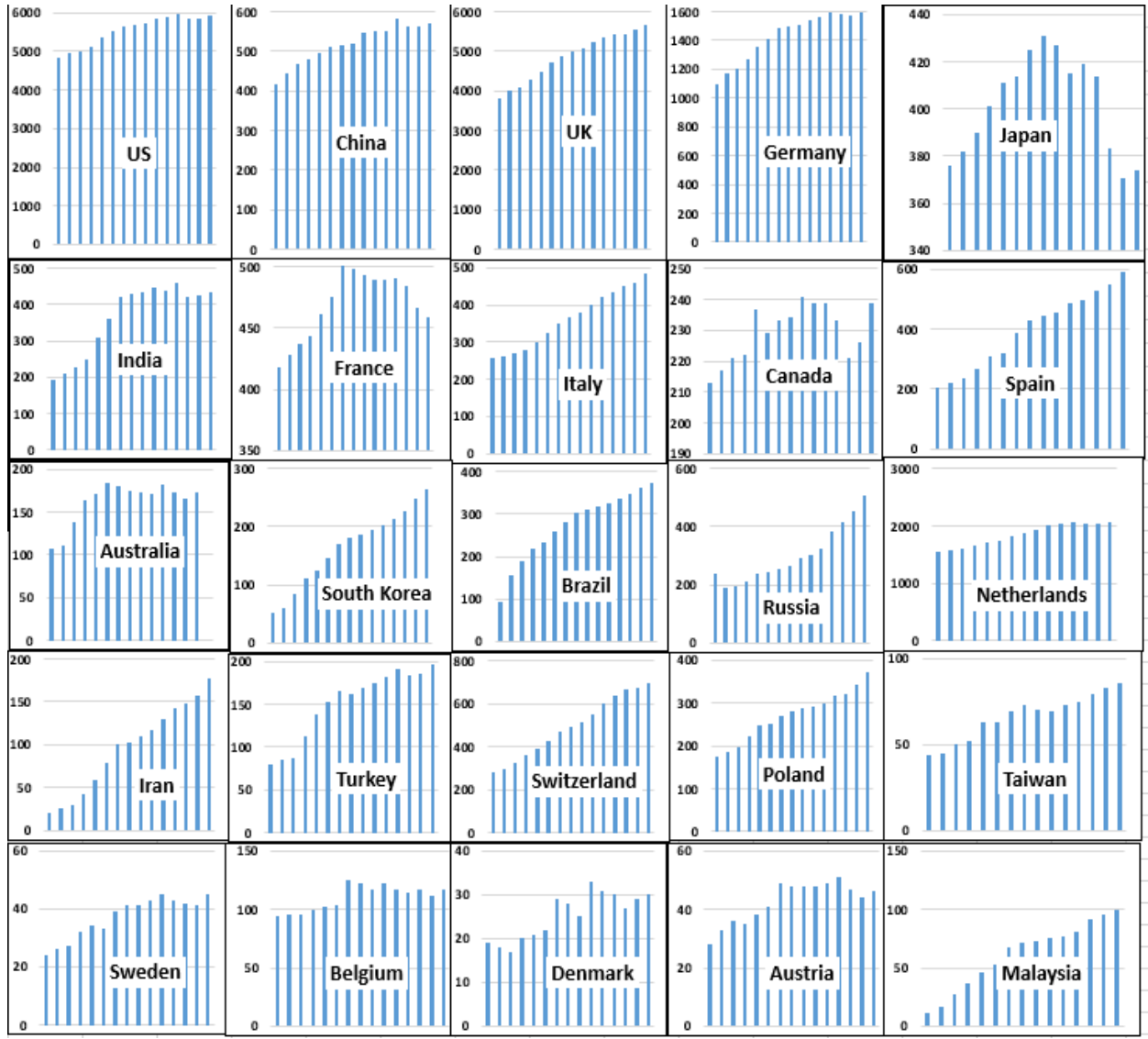

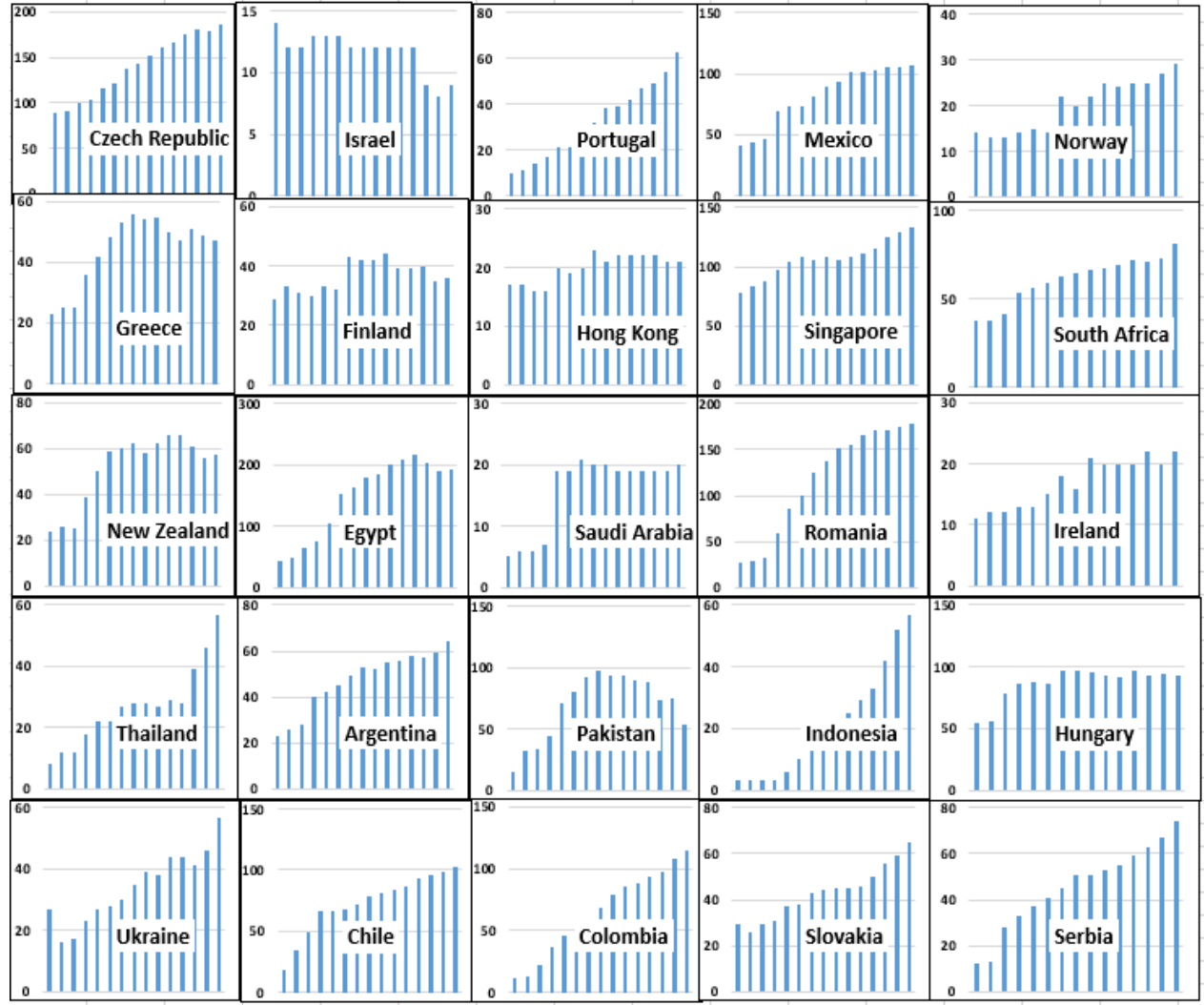

**Figure 2:** Number of journals (NCJ) published from a country during 2005-19 for the 50 countries. (x-axis shows the year and y-axis is the number of journals. Each bar represents the number of journals published from a country in a specific year.)

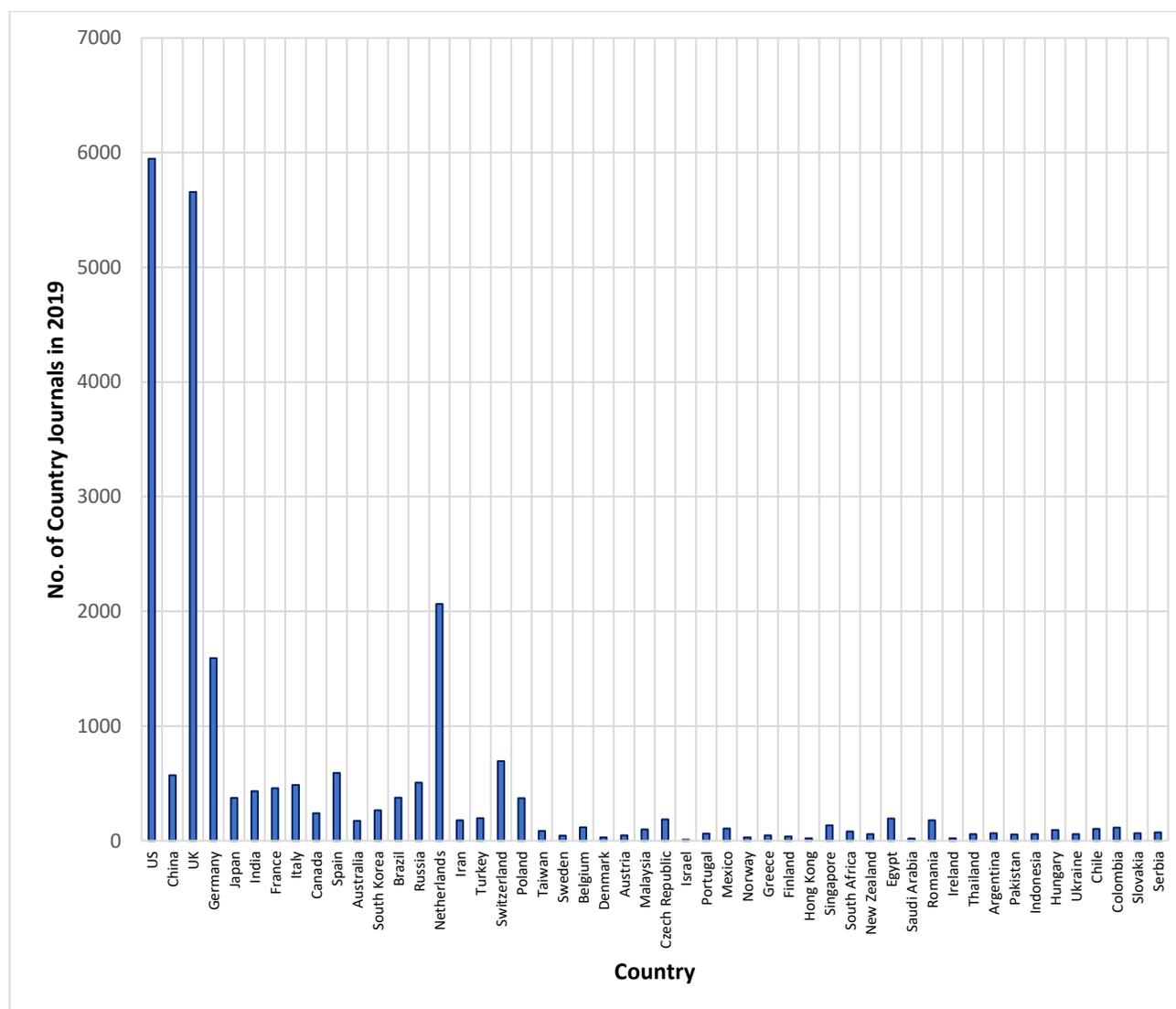

**Figure 3:** Number of journals (NCJ) published by different countries in the year 2019. (x-axis represents countries and y-axis represents number of journals. Each bar represents the number of journals published from a country.)

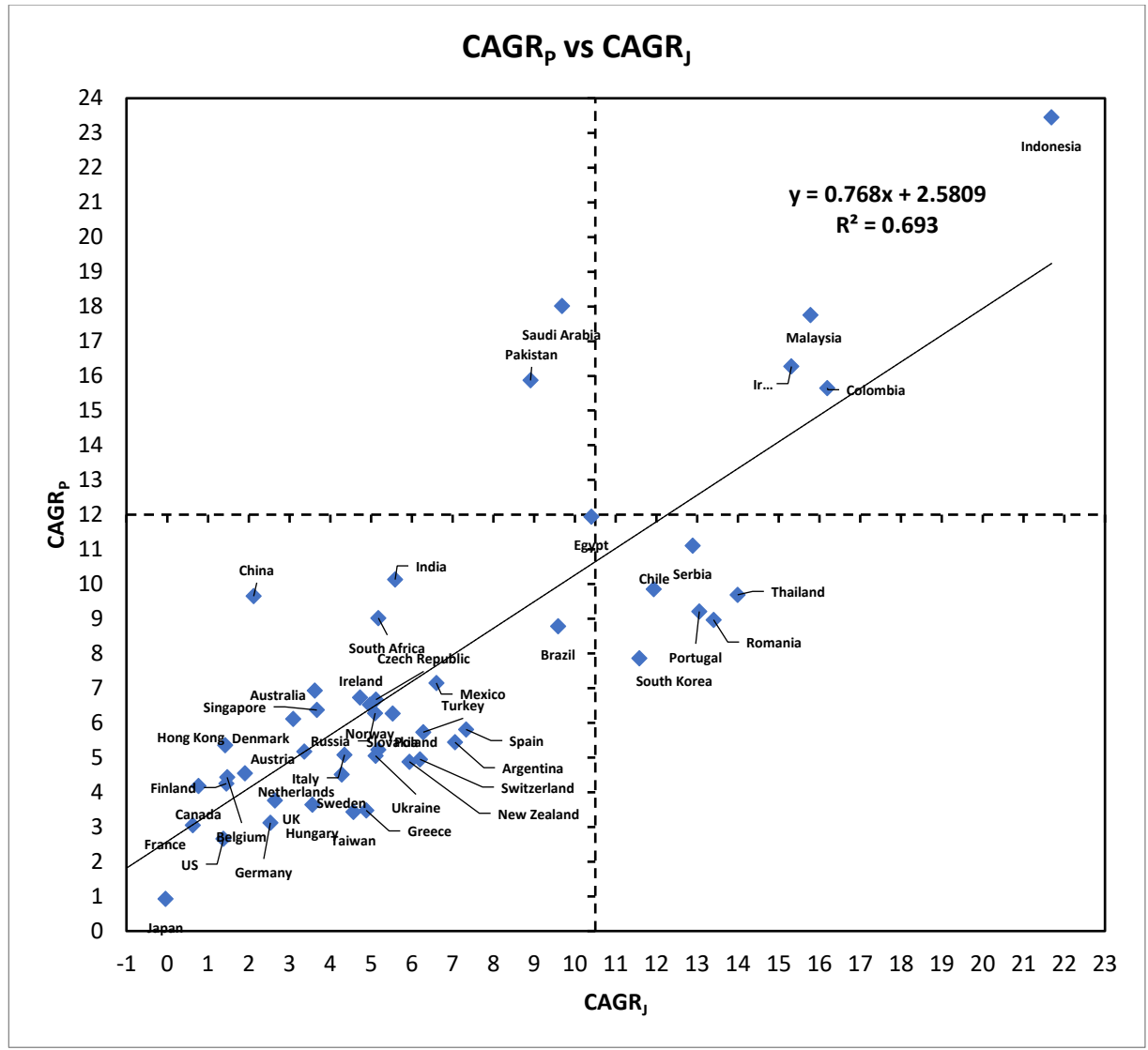

**Figure 4:** $CAGR_J$ vs $CAGR_P$. (Each (x,y) point represents a country.)

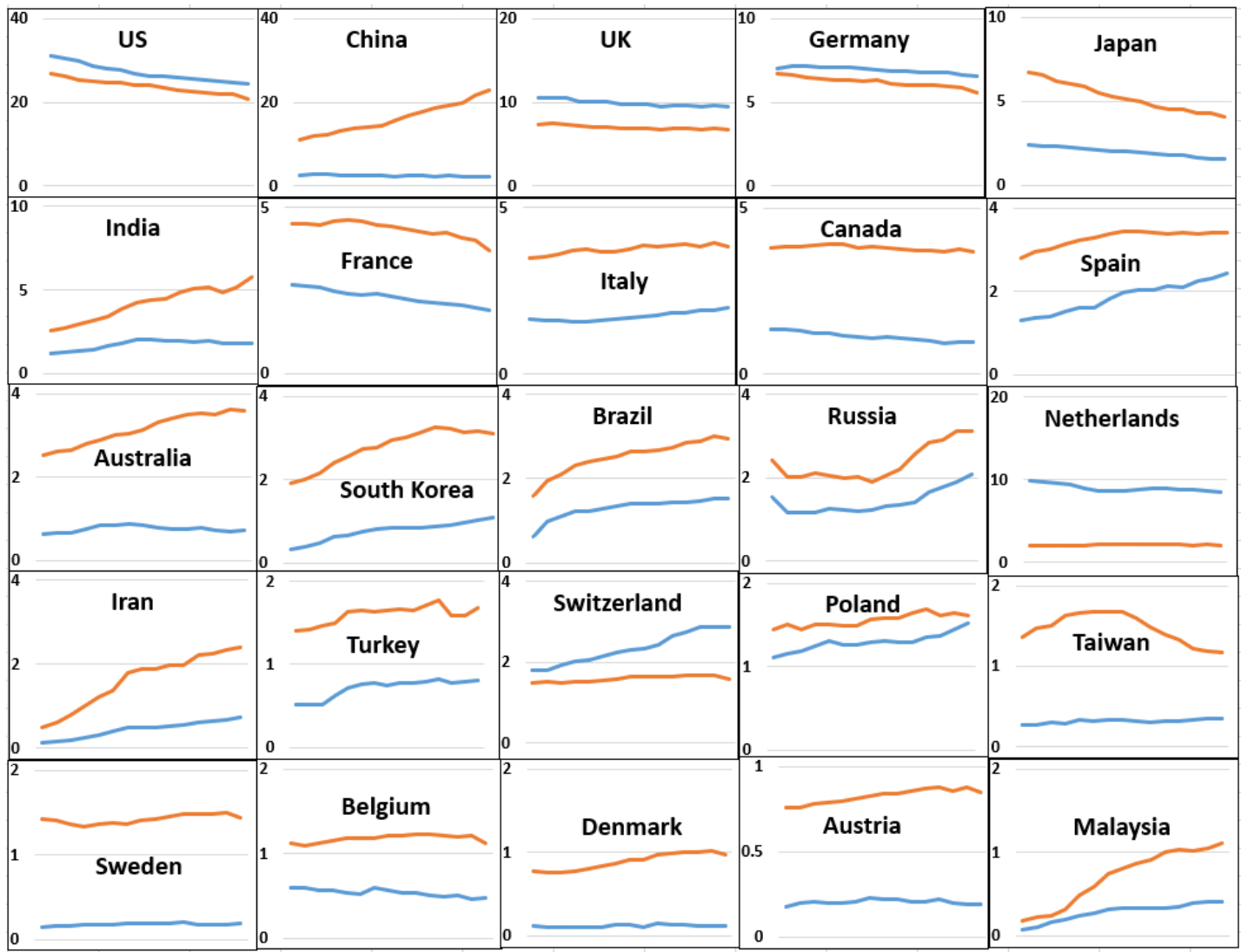

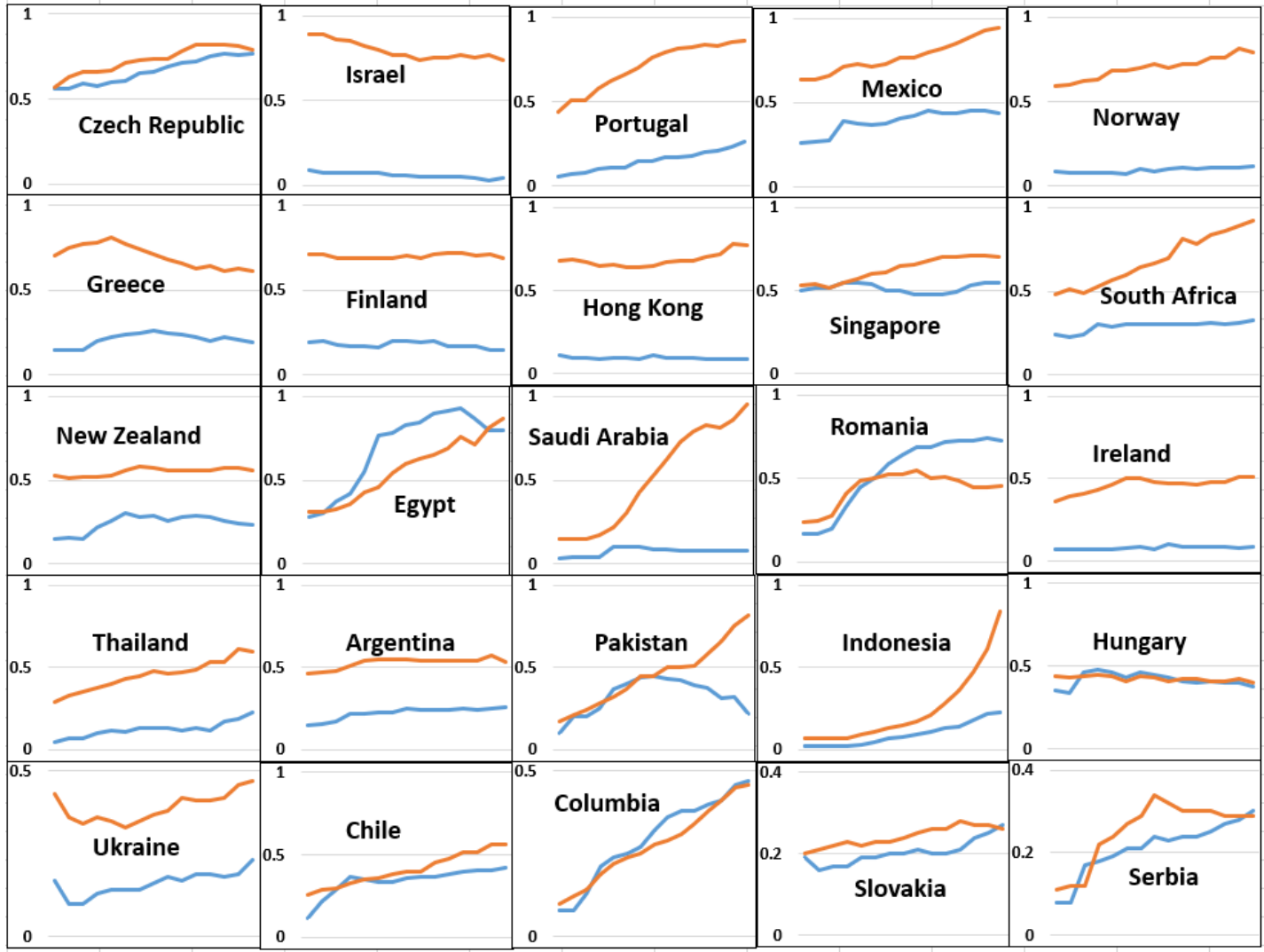

Figure 5: X1 (blue line) and X2 (red line) with respect to time. (The x-axis represents year and the y-axis represents value. X1-ratio of country journals to world-wide journals, and X2- ratio of country papers to world-wide papers)

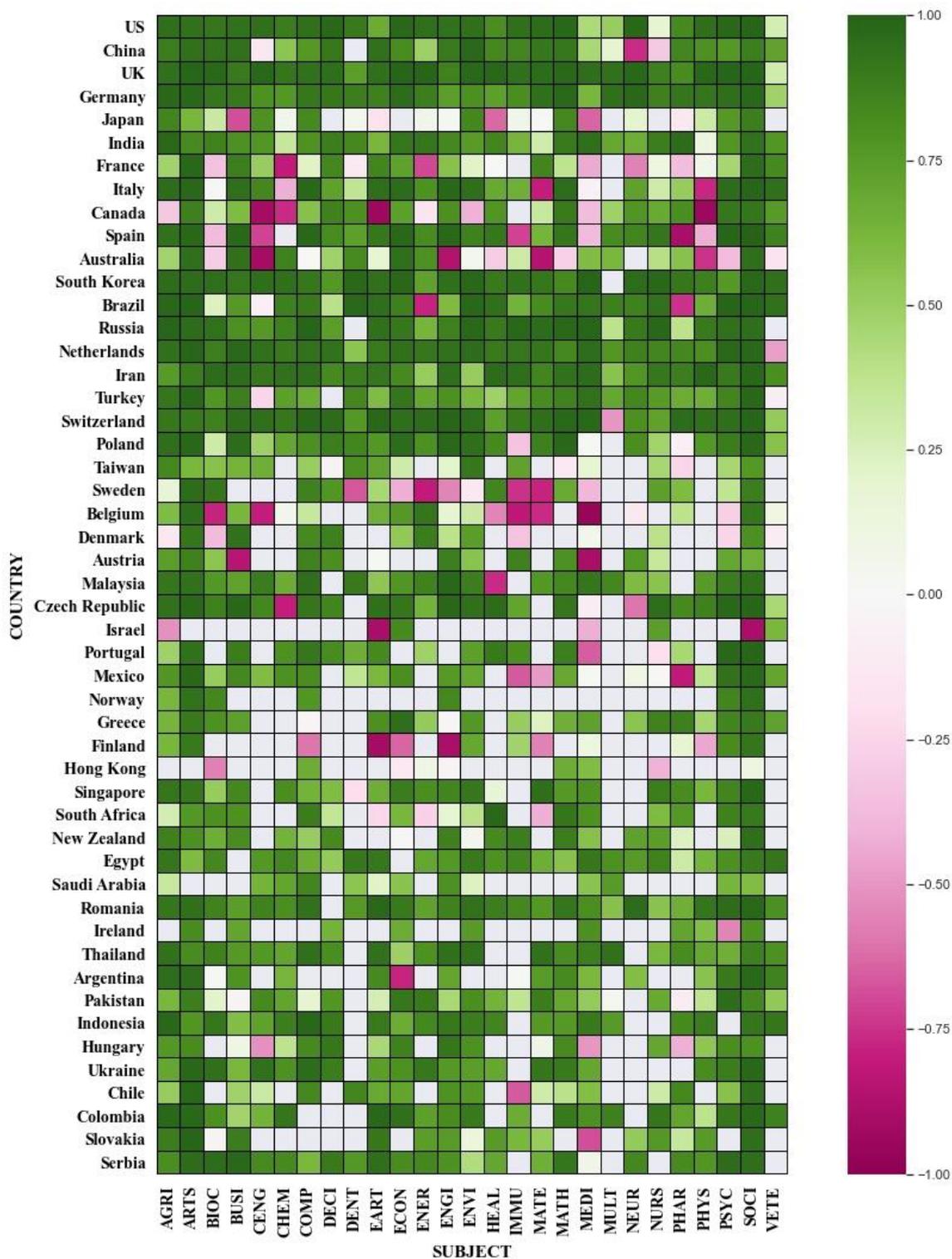

**Figure 6:** Correlations between NCJ-S and TP-S for different countries and subjects.

**Note:** **AGRI**: Agricultural and Biological Sciences; **ARTS**: Arts and Humanities; **BIOC**: Biochemistry, Genetics and Molecular Biology; **BUSI**: Business, Management and Accounting; **CENG**: Chemical Engineering; **CHEM**: Chemistry; **COMP**: Computer Science; **DECI**: Decision Sciences; **DENT**: Dentistry; **EART**: Earth and Planetary Sciences; **ECON**: Economics, Econometrics and Finance; **ENER**: Energy; **ENVI**: Environmental Science; **HEAL**: Health Professions; **IMMU**: Immunology and Microbiology; **MATE**: Materials Science; **MATH**: Mathematics; **MEDI**: Medicine; **MULT**: Multidisciplinary; **NEUR**: Neuroscience; **NURS**: Nursing; **PHAR**: Pharmocology, Toxicology and Pharmaceutics; **PHYS**: Physics and Astronomy; **PSYC**: Psychology; **SOCI**: Social Sciences; **VETE**: Veterinary.